\documentclass[aps,pre,showpacs,showkeys,twocolumn,superscriptaddress,
floats,floatfix,preprintnumbers]{revtex4-1}
\usepackage{amssymb,amsmath}
\usepackage{bm,url,graphics,subfigure,epsfig,rotating,float}
\usepackage{mathrsfs} 
\usepackage{color,soul}
\usepackage{ulem}
\graphicspath{{./}{fig/}}
\def \tN {{\tt N}}
\def \mGQ  {\mathcal{G}_{\rm Q}}
\def \mAQ  {\mathcal{A}_{\rm Q}}
\def \SQ  {\mathcal{S}_{\rm Q}}
\def \EQ  {\mathcal{E}_{\rm Q}}
\def \PiQ  {\Pi_{\rm Q}}
\def \DQ   {D_{\rm Q}}
\def \SigmaQ   {\Sigma_{\rm Q}}
\def \qstar {q_{\ast}}
\def \bE  {\mathbb{E}}
\def \bS  {\mathbb{S}}
\def \tA {{\tt A}}

\def \qq {\bm{q}}
\def \phat {\hat{\phi}}
\def \vhat {\hat{{\bm v}}}

\def \JJ {\bm{J}}
\def \mE {\mathcal{E}}

\def \mL   {\mathcal{L}}
\def \Phiz {\Phi_{\rm 0}}
\def \mF {\mathcal{F}}

\def \uu  {{\bm u}}

\def \vv  {{\bm v}}

\def  \xx  {{\bm x}}

\def \grad {{\bm \nabla}}

\def \dive {{\bm \nabla}\cdot}
\def \lap {\nabla^2}
\def \delt {\partial_t}

\newcommand{\bra}[1]{\left\langle #1\right\rangle}
\def \Pe {\mbox{Pe}}
\def \Ch {\mbox{Ch}}
\def \La {\mbox{La}}

\def \Sc {\mbox{Sc}}

\def \Rey  {\mbox{Re}}

\def\i{{\rm i}}

\newcommand{\eq}[1]{~(\ref{#1})}
\newcommand{\Eq}[1]{Eq.~(\ref{#1})}
\newcommand{\Fig}[1]{Fig.~(\ref{#1})}
\newcommand{\subfig}[2]{Fig.~(\ref{#1}#2)}

\newcommand{\bfig}{\begin{figure}}
\newcommand{\efig}{\end{figure}}
\newcommand{\bc}{\begin{center}}
\newcommand{\ec}{\end{center}}
\newcommand{\bea}{\begin{eqnarray}}
\newcommand{\eea}{\end{eqnarray}}

\begin{document} 
\title{Coagulation drives turbulence in binary fluid mixtures } 
\author{Akshay Bhatnagar}
\email{akshayphy@gmail.com}
\affiliation{ Nordita, KTH Royal Institute of Technology and
Stockholm University, Roslagstullsbacken 23, 10691 Stockholm, Sweden}
\author{Prasad Perlekar}
\email{perlekar@tifrh.res.in}
\affiliation{TIFR Centre for Interdisciplinary Sciences,
  Tata Institute of Fundamental Research,  Gopanpally, Hyderabad 500046,
  India.}
\author{Dhrubaditya Mitra}
\email{dhruba.mitra@gmail.com}
\affiliation{ Nordita, KTH Royal Institute of Technology and
Stockholm University, Roslagstullsbacken 23, 10691 Stockholm, Sweden}

\begin{abstract}
  We use direct numerical simulations and scaling arguments to study 
  coarsening in binary fluid mixtures with a conserved order parameter in
  the droplet-spinodal regime -- the volume fraction of the droplets is
  neither too small nor symmetric -- for small diffusivity and viscosity.
  Coagulation of droplets drives a turbulent flow that eventually decays. 
  We uncover a novel coarsening mechanism, driven by turbulence where the 
  characteristic length scale of the flow is different from the characteristic
  length scale of droplets, giving rise to a domain growth law of
  $t^{1/2}$, where $t$ is time.
  At intermediate times, both the flow and the droplets form self-similar
  structures:
  the structure factor $\bS(q) \sim q^{-2}$ and the kinetic energy
  spectra $\bE(q) \sim q^{-5/3}$ for an intermediate range of $q$,
  the wavenumber. 
\end{abstract}
\preprint{NORDITA 2021-081}
\maketitle
\begin{figure*}
  \includegraphics[width=0.24\textwidth]{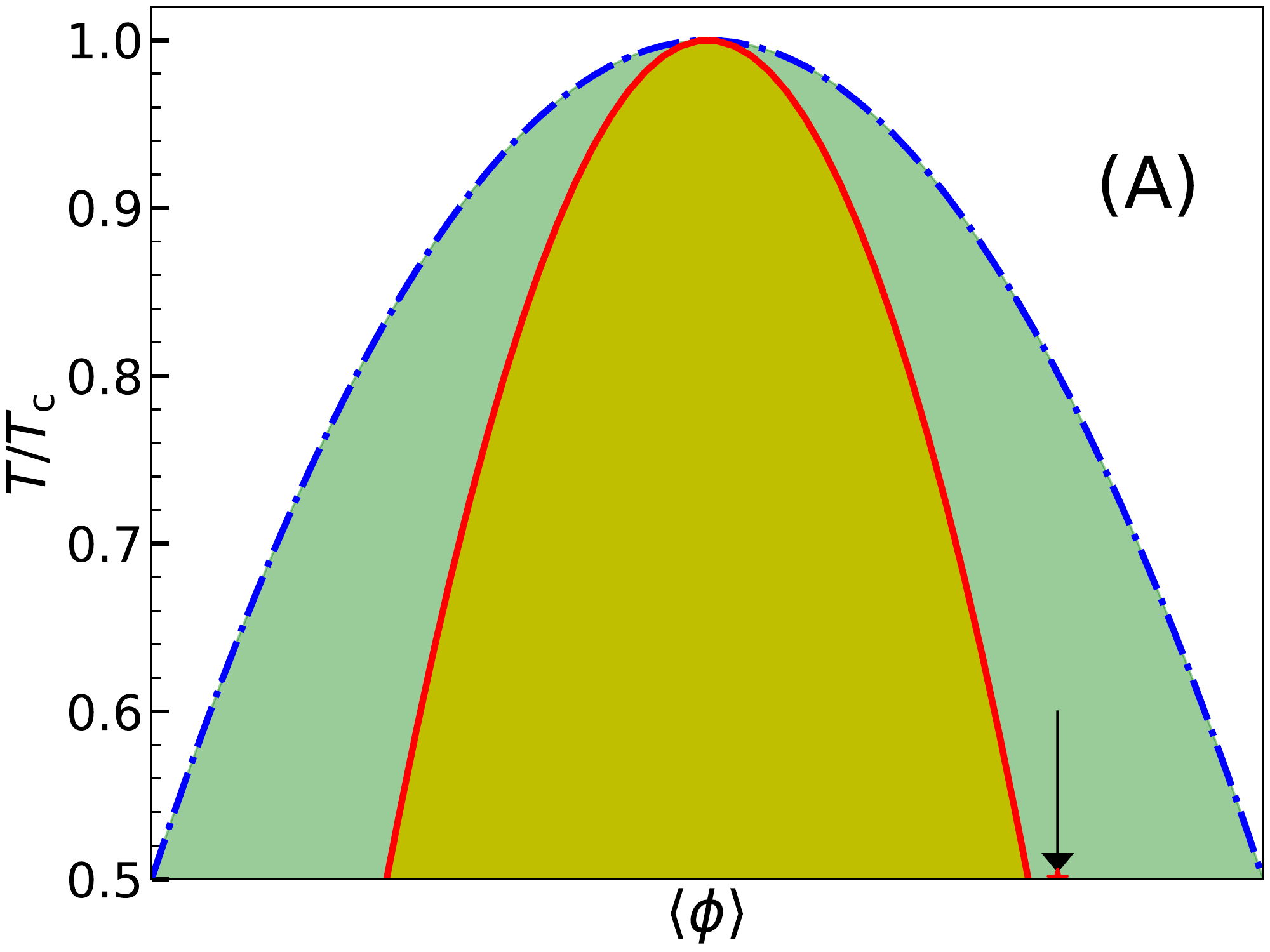}
  \includegraphics[width=0.24\textwidth]{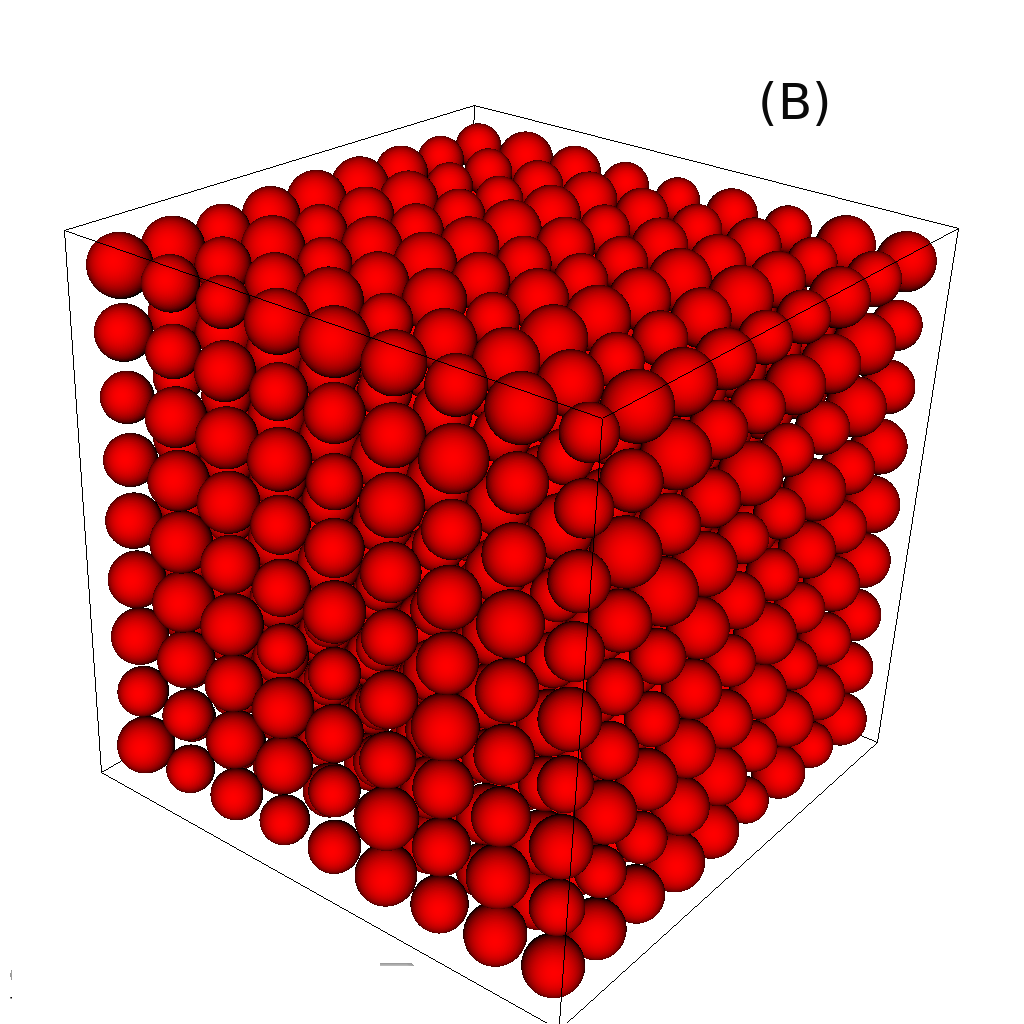}
  \includegraphics[width=0.24\textwidth]{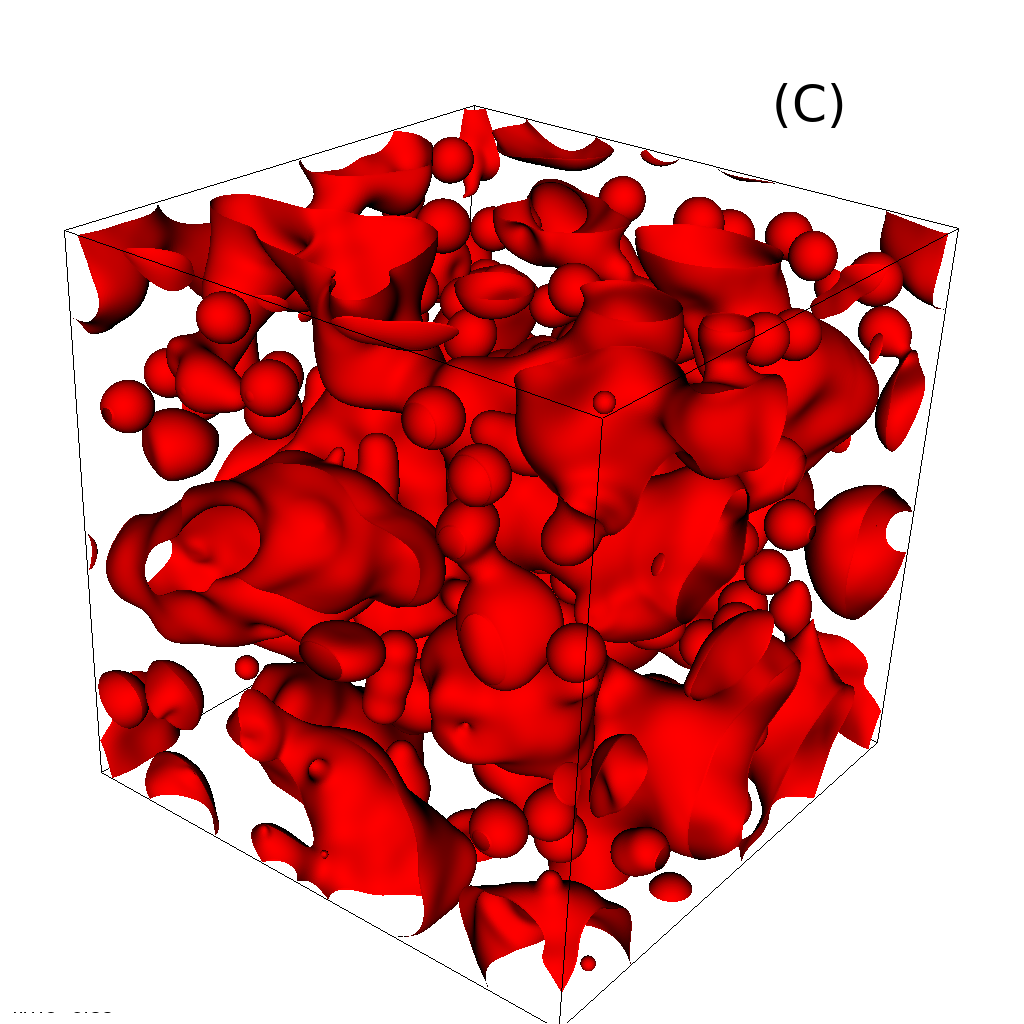}
  \includegraphics[width=0.24\textwidth]{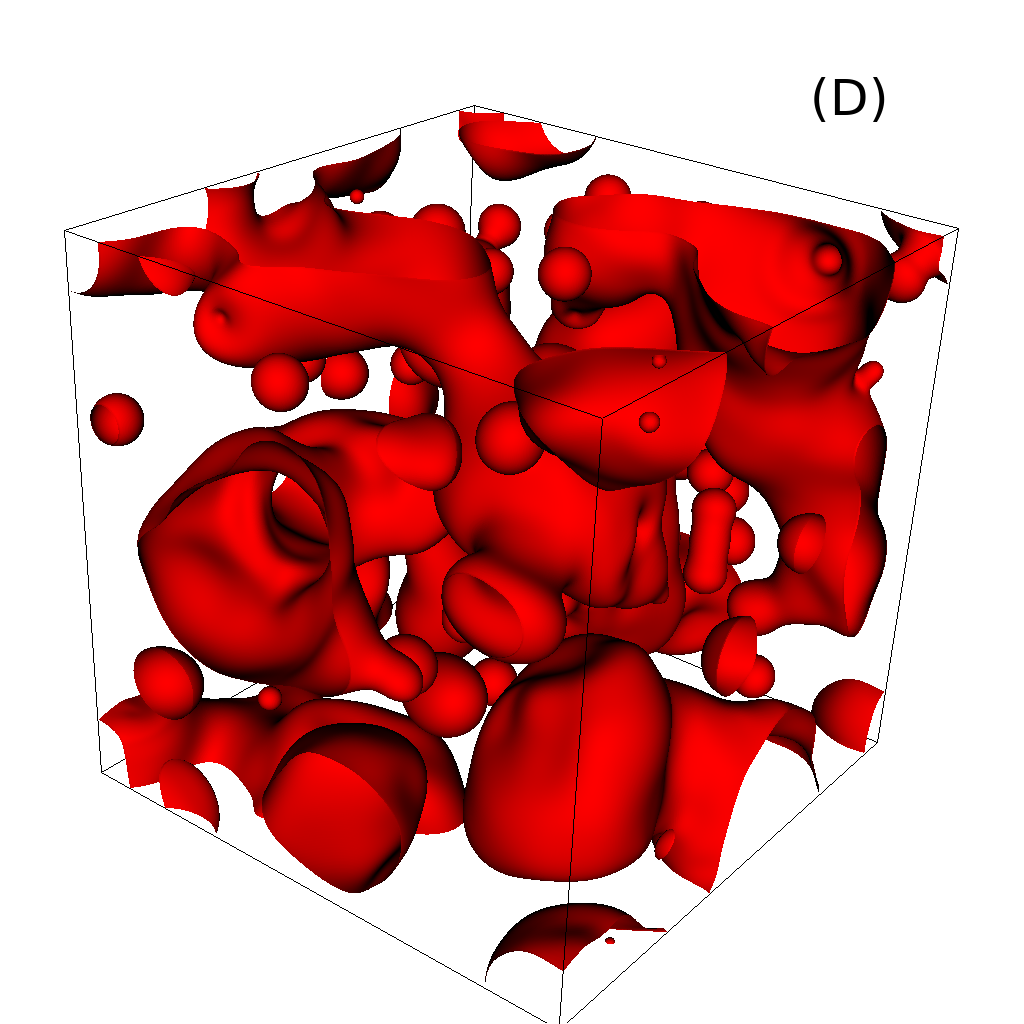}
\caption{\label{fig:pdiagram} (A) A sketch of the phase-diagram
of Landau-Ginzburg type $\phi^4$ theory. The blue dashed line is
the phase coexistence curve. The red line is the spinodal curve.
Near the phase coexistence line droplets form by nucleation.
Inside the red line domains grow by spinodal decomposition.
The point marked by a red star,
pointed at by the arrow, shows our position in this phase-diagram;
this is outside the spinodal curve but not too close to the phase
coexistence line -- droplet spinodal decomposition operates here.
(B), (C), (D): Isosurfaces of $\phi$ at $\phi=0$ for 
$\La = 10^5$ at times $\tau = 0$, $18$, and $35$ respectively. 
}
\end{figure*}

The non-equilibrium dynamics of phase separation plays a crucial role in
many different branches of physics, e.g., in condensed matter
systems~\cite{hoh+hal77, Cha+Lub98,cates2000soft,bray2002theory,Onuki2002phase,
cates2017complex, cates2018theories} both classical and
quantum~\cite{hoffer1986dynamics,hofmann2014coarsening,
  mendoncca2012photon}, nuclear matter~\cite{chomaz2004nuclear},
cosmology~\cite{boyanovsky2006phase}, and astrophysics~\cite{
  prendergast1973photon,rieutord2006phenomenological}. 
To set the scene, consider 
the canonical model of equilibrium phase transition: 
the Landau-Ginzburg type $\phi^4$ theory
with a scalar order-parameter $\phi$~\cite{Cha+Lub98}. 
We show a sketch of its equilibrium phase diagram in
\subfig{fig:pdiagram}{A}.
When the system is quenched from an  uniform
high-temperature phase to a state below the coexistence
curve the uniform phase is no longer in stable thermal equilibrium.
The system approaches equilibrium -- two co-existing domains
with $\phi=1$ and $\phi=-1$ separated by a domain wall
-- by phase separating.
We consider the case where the order parameter is conserved -- 
\textit{model B} of Hohenberg and Halperin~\cite{hoh+hal77}.
If the thermal noise is ignored -- which is the case
in the rest of this paper --  model B reduces to the Cahn-Hilliard
equation~\cite{cahn1958free}. 
Domain growth in the Cahn-Hilliard equations shows a
variety of dynamical behavior that has been uncovered by
analytical and numerical
techniques~\cite[see e.g.,][]{Cha+Lub98,Onuki2002phase,bray2002theory,
  leyvraz2003scaling}. 

In binary fluid mixtures, e.g., oil-water systems, flows are also
coupled with the phase separation dynamics --
the Cahn--Hilliard--Navier--Stokes equations or the \textit{model H} of Hohenberg and
Halperin without noise.
There are a plethora of possible growth mechanisms and
corresponding growth laws~\cite{cates2018theories}
that have been investigated:
(a) diffusive growth, essentially by the
Lifshitz--Slyozov--Wagner mechanism~\cite{lifshitz1961kinetics,wagner1961theory}
that operates in model B;
(b) collisional growth due to Brownian motion of the
droplets~\cite{binder1974theory,siggia1979late,san1985phase,tanaka1996coarsening};
(c) growth due to the viscous flows \cite{siggia1979late}; and
(d) growth due to inertial flows~\cite{furukawa1985effect, alexander1993hydrodynamic,
  kendon2001inertial}.
The mechanisms  (a) and (b) give the growth law $L \sim t^{1/3}$;
(c) gives $L \sim t$; and (d)  $ L \sim t^{2/3}$.
Furthermore coarsening also depend on initial condition --
whether it is a critical quench (inside the spinodal curve) or an off-critical
quench (near the phase-coexistence curve)~\cite{farrell1991growth,
  datt2015morphological, shimizu2015novel}.
The possible growth mechanisms and the  growth laws have been extensively
studied experimentally too~\cite{chou1979phase,wong1981light,perrot1994nucleation,
  rahman2019viscous}.
Dynamics of domain growth near the co-existence line and
well inside the spinodal for not too small diffusivity and viscosity
are reasonably well understood. 

Due to the necessity of massive computational resources,
the part of parameter space with  small diffusivity and viscosity remained unexplored. 
Recently, Naso and Nar\'aigh~\cite{naso2018flow}, for the first time,  obtained
signatures of novel coarsening behavior in this regime.
In this paper, we lay bare the growth-laws and the 
turbulent flows that develop in this regime using scaling theory and analysis of
the largest
direct numerical simulations of the Cahn--Hilliard--Navier--Stokes equations.
In particular, we show that if the system is initialised in the
droplet-spinodal regime~\cite{shimizu2015novel} coagulation of droplets drive a
nonlinear flux of kinetic energy that gives rise to Kolmogorov-like turbulence
and a scaling of $L \sim t^{1/2}$ at intermediate times.
We also present a scaling theory of this phenomena. 

The Cahn--Hilliard--Navier--Stokes equations are given by:
\begin{subequations}
  \begin{align}
    \delt\phi + \dive\JJ &= 0 \quad\/.\label{eq:phit} \\
    \JJ &= \vv\phi - \Gamma \grad \mu \quad \/.\label{eq:flux}\\
    \mu &= \frac{\delta \mF}{\delta \phi} \quad\/.\label{eq:mu}\\
    \mF[\phi] &= \frac{\Lambda}{\xi^2}\int d^dx\left[ f(\phi)
               + \frac{\xi^2}{2}\mid \grad\phi \mid^2 \right] \quad\/.\label{eq:LG}\\
    f(\phi) &= \frac{1}{4}\left(1 - \phi^2 \right)^2 \quad\/.\label{eq:fenergy}\\
  \rho(\delt + \vv\cdot\grad)\vv &= \eta\lap\vv - \rho\grad p -\phi\grad\mu\/.
                      \label{eq:mom}\\
    \dive \vv &=0 \/.\label{eq:divv}
  \end{align}
  \label{eq:CH}
\end{subequations}
Here $\phi$ is the order-parameter,
$\vv$ the velocity of the flow,
$\mu$ the chemical potential,
$\Gamma$ the transport coefficient of the chemical potential,
$\eta = \rho\nu$ is the dynamic viscosity, $\rho$ is the density,
$\nu$ the kinematic viscosity,
and $\xi$ is the length-scale
the characterizes the interface thickness.
The surface tension
$\sigma = (2\sqrt{2}/3)\Lambda/\xi$

We use the initial size of the droplets $R$ as the characteristic
length scale and  $V = \nu/R$ as the characteristic velocity.
The non-dimensional parameters are 
the Laplace number $\La \equiv \sigma R/(\rho\nu^2)$,
the Schmidt number $\Sc \equiv \nu/D$, and
the Cahn number $\Ch \equiv \xi/R$, 
where the diffusivity $D \equiv \Gamma\Lambda/\xi^2$.
In all our simulations $\Sc=1$ and we
use several Laplace numbers and two different Cahn numbers.
We use a spectral code with periodic boundary conditions
with $\tN^3$ resolutions where $\tN = 512$ and $1024$ --
the highest resolution simulations done for this 
system~\footnote{
 A comprehensive description of the algorithm, the complete 
 list of parameters and the non-dimensional equations are 
 given in Appendix \ref{sec:MM}.
 A comparison of our parameters with Ref.~\cite{naso2018flow} and
 Ref.~\cite{kendon2001inertial} is given in Appendix \ref{sec:comparison}.
}.

We choose all the droplets to have the same radius $R/\mL=1/(8\pi)$
and their centers placed on a cubic lattice.
We then add small random perturbations to the radii of
the droplets, see \subfig{fig:pdiagram}{B}. 
The initial volume fraction occupied by the
droplets (minority phase) is approximately $0.2$ such that
$\bra{\phi} = 0.62$, where $\bra{\cdot}$ denotes spatial averaging. 
This choice of $\bra{\phi}$,
marked in the phase-diagram, \subfig{fig:pdiagram}{A},
puts us in the droplet spinodal decomposition
regime~\cite{shimizu2015novel} --  we are neither well
inside the spinodal curve nor very close to
the co-existence curve.

We show how coarsening progresses in \subfig{fig:pdiagram}{B} to 
\subfig{fig:pdiagram}{D}. 
At very early times the drops remains practically unchanged in size
but move -- typically towards their closest neighbour.
At $t = 0$ there was no flow -- the flows that move the
droplets are generated by compositional Marangoni
effect~\cite{shimizu2015novel}.
To confirm, we simulate a collection of seven drops - one central drop
at the origin and six drops placed on a cubic lattice around it.
Due to the asymmetry all the peripheral drops move towards the
central drop and eventually merge into one.
A similar experiment where we place the drops on a line also show that
drops move toward their nearest neighbours. 

We define the structure factor and the energy spectrum as
the shell-integrated Fourier spectra of $\phi$ and $\vv$
respectively, i.e., 
\begin{subequations}
  \begin{align}
    \bS(q) &\equiv \int \phat(\qq)\phat(-\qq)d\Omega\quad\/, 
      \label{eq:Sq} \\
      \bE(q) &\equiv \int \vhat(\qq)\vhat(-\qq)d\Omega\quad\/,
      \label{eq:Eq}
  \end{align}
\end{subequations}
where $\phat(\qq)$ and $\vhat(\qq)$ are respectively the
Fourier transforms of $\phi(\xx)$ and $\vv(\xx)$,
$q = \mid \qq \mid$ is the magnitude of the wavevector $\qq$,
and $\Omega$ is the solid angle in Fourier space. 
We calculate the evolving, characteristic length
scale, $L$ as~\cite{perlekar2014spinodal,shimizu2015novel}
\begin{equation}
K \equiv \frac{\int dq q \bS(q) }{\int dq \bS(q)} \/\quad
{\rm and}\quad L \equiv 2\pi/K
\label{eq:kandl}
\end{equation}

\begin{figure}
  \includegraphics[width=0.95\columnwidth]{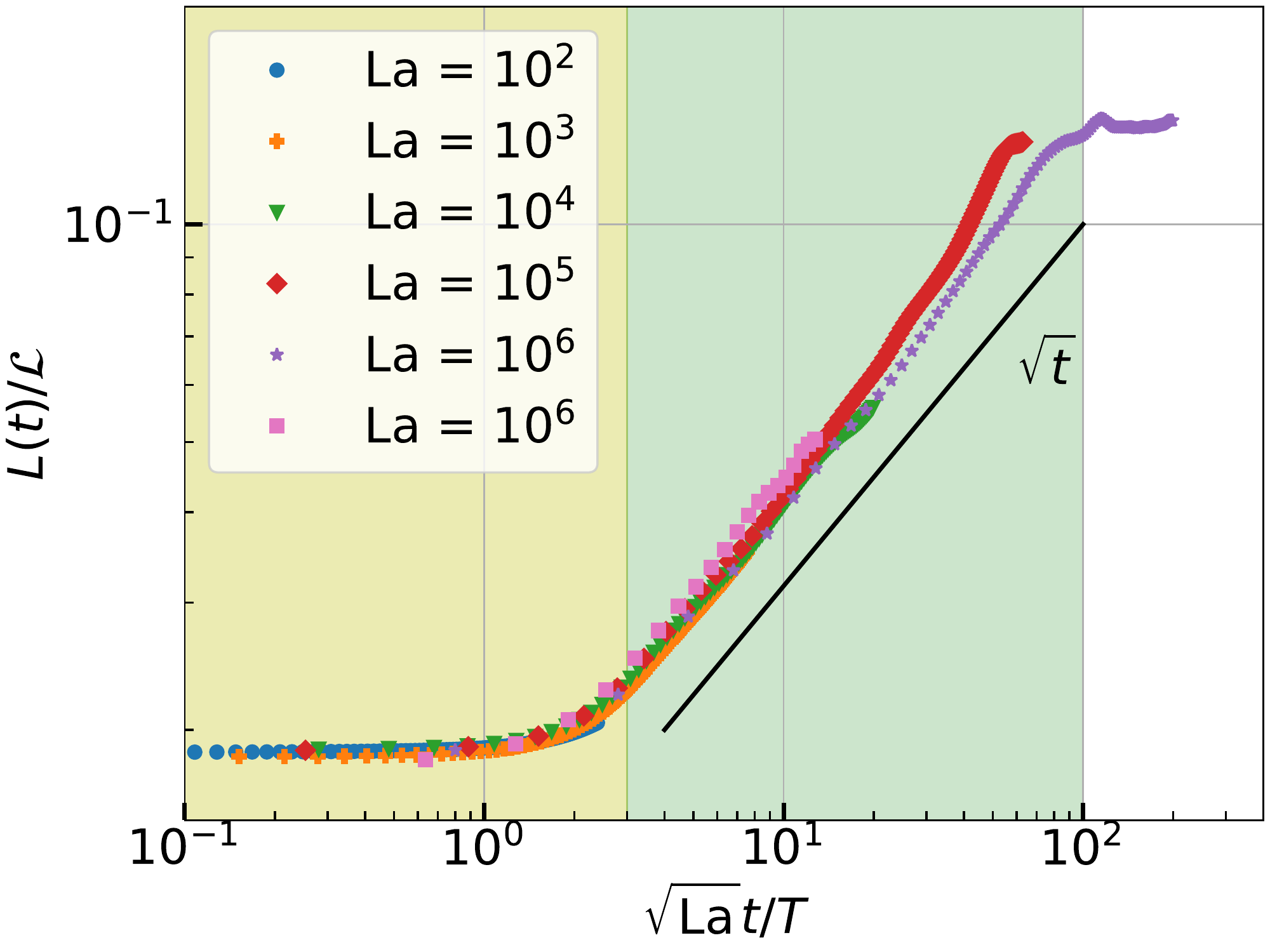}\\
  \includegraphics[width=0.95\columnwidth]{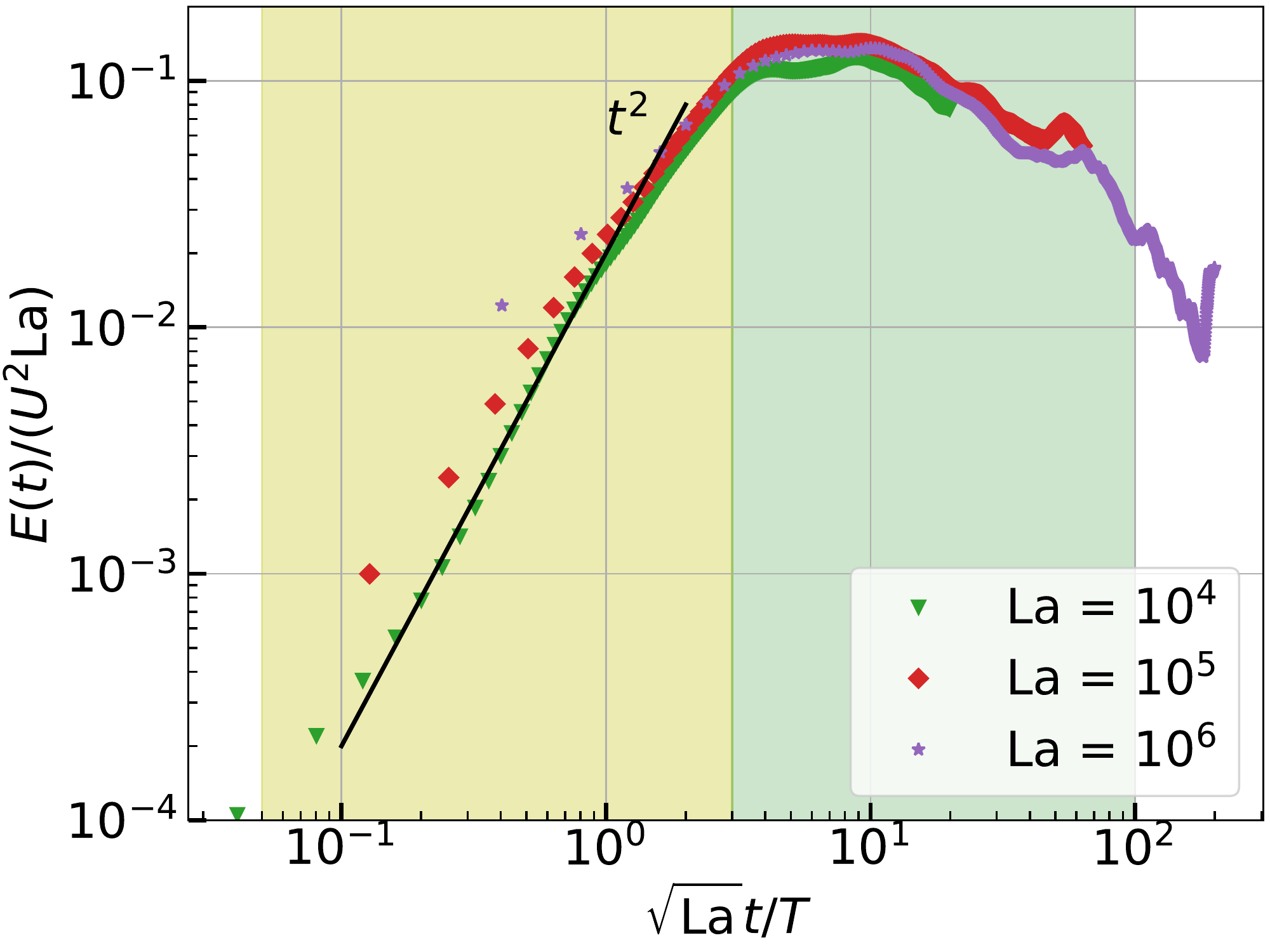}
  \caption{\label{fig:growth} The evolution of the 
    (A) The characteristic length-scale, $L(t)$, and
    (B) the total kinetic energy $E(t)$ for all the $\tA$ runs, $\Sc=1$. 
    The data for different Laplace number collapse on each other
    when plotted as a function of the scaled time
    $\tau \equiv \sqrt{\La}t/T$.}   
\end{figure}
The time evolution of the characteristic length scale $L(t)$ and 
the total kinetic energy $E(t)$ are shown in \Fig{fig:growth}.
The evolution for different Laplace numbers collapse when plotted
as a function of scaled dimensionless time
$\tau \equiv \sqrt{\La}(t/T)$.
At short times $\tau \lesssim 3$, $L$ is practically a constant
-- very few droplets have merged.
During the same time interval the kinetic energy of the flow
grows as $E(t) \sim t^2$. 
At intermediate times $10 \lesssim \tau \lesssim 100$
the kinetic energy of the flow remains almost a constant
(decreases slowly) while the characteristic length scale
grows as $L(t) \sim t^{1/2}$ for at least a decade. 
We also detect a dependece on Cahn number -- the transition to
$L\sim t^{1/2}$ happens earlier for larger Cahn number. 
At very late times $100 \lesssim \tau$, as expected, the kinetic energy
starts to decay fast while $L$ saturates.  
The scaling of $L$ with a dynamic exponent of $1/2$
has been observed before in two-dimensional
simulations~\cite{wu1995effects} in the presence of noise
and for an off-critical quench 
but never before in three dimensions~\footnote{
  Ref~\cite{osborn1995lattice} has found the exponent $1/2$ for
  liquid-gas systems, which is similar to model A of Hohenberg and Halperin,
  but not for binary fluids.
  Ref.~\cite{wagner1998breakdown} also obtained $1/2$ but for a length
  scale that is different from $L$ for a case with low viscosity where
  they found that general scale invariance is broken -- length scales defined
  in different ways scale differently.
  Both of these simulations are in two dimensions.}.

\begin{figure}
  \includegraphics[width=0.95\columnwidth]{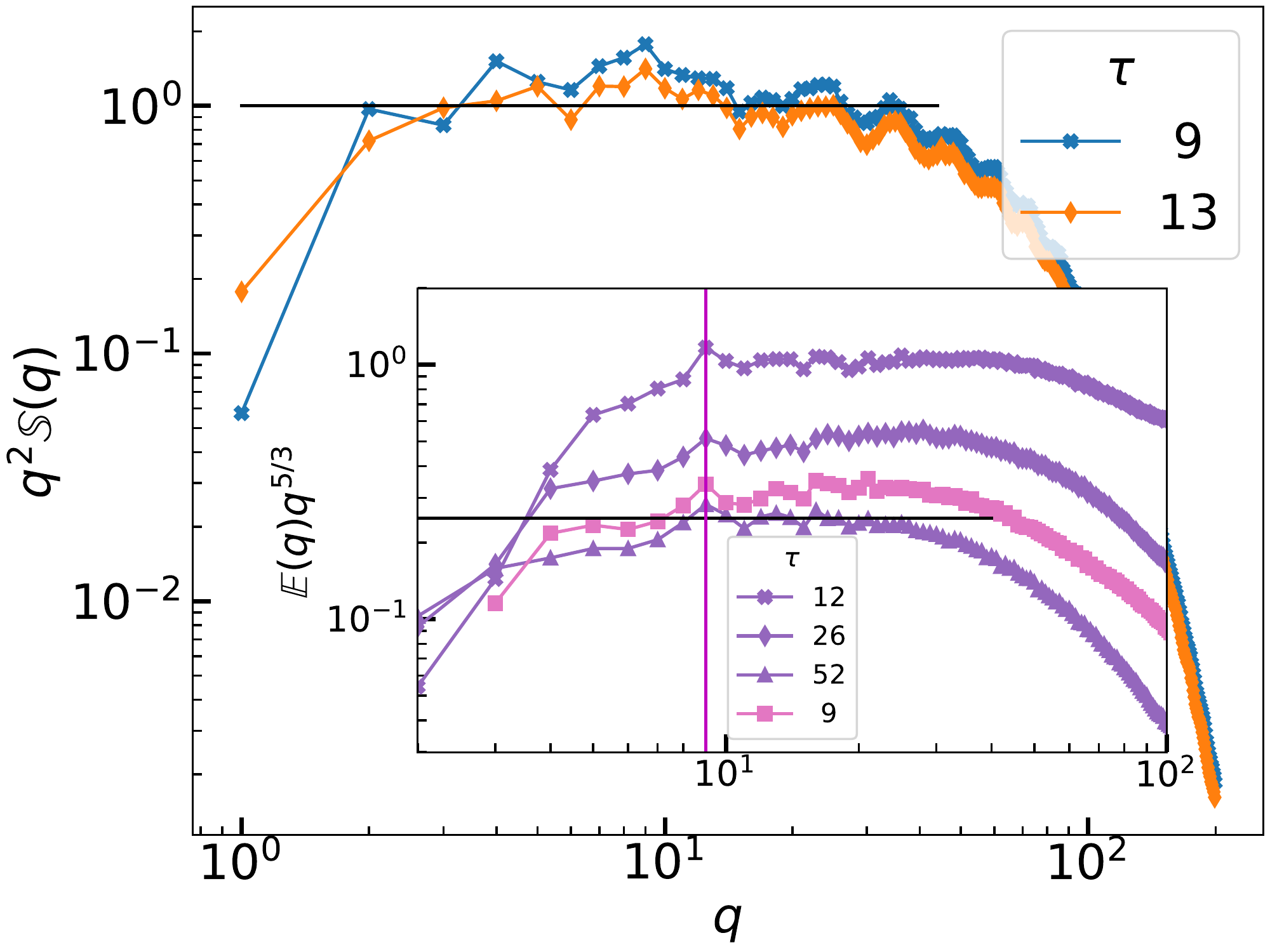}
  \caption{\label{fig:specu} The compensated structure factor,
    $q^2\bS(q)$ as a function of $q$ in log-log scale at two different
    times for the run {\tt A6} ($\tN = 1024$).
    A range of $q$ over which the compensated plot is flat
    implies $\bS(q) \sim q^{-2}$ over that range.
    We plot a horizontal line for comparison.
    Inset: The compensated kinetic energy spectra,
    $q^{5/3}\bE(q)$ as a function of $q$ in log-log scale at three different
    times for the run {\tt A3}($\tN=512$) and at $\tau=9$ for the
    run {\tt A6} ($\tN = 1024$)
    A range of $q$ over which the compensated spectra is flat
    implies $\bE(q) \sim q^{-5/3}$ over that range.
    We plot a horizontal line for comparison.
    The compensated spectra has a peak at $q = q_{\ast}\approx 9$ shown
    by a vertical line. Note that $q_{\ast}$ does not change with time. 
  }
\end{figure}

In \Fig{fig:specu} we show representative plots of the
\textit{compensated} structure factor, $q^2\bS(q)$ and the
\textit{compensated} energy spectrum, $q^{5/3}\bE(q)$
for times during which the $L\sim t^{1/2}$ scaling is observed.
In both cases we find a region in $q$ over which the
compensated plots are approximately horizontal.
This implies that over this range of $q$,
\begin{equation}
  \bS(q) \sim q^{-2} \quad\text{and}\quad \bE(q) \sim q^{-5/3}
  \label{eq:spec}
\end{equation}
 The change in $\phi$ across a length scale $r$,
$\delta_r\phi$ has two contributions, one
smooth -- if $r$ lies either well inside a droplet or well outside droplets --
and the another independent of $r$ -- if $r$ lies across the boundary of
a droplet. In other words, there are isolated jumps connected by smooth
regions which implies~\cite[section 8.5.2]{Fri96}, similar to Burgers
equation~\cite{aurell1992multifractal}, $\bS(q) \sim q^{-2}$. 
The scaling $\bE(q) \sim q^{-5/3}$ signifies fully developed
turbulence a-la Kolmogorov.
This plays a key role in the scaling theory we discuss next. 

We construct a scaling theory by generalizing the
standard approach~\cite{Cha+Lub98}.
From \Eq{eq:phit} the rate of change of characteristic
length scale of a droplet is given by
$\phi dL/dt \sim J$ where $J \equiv \mid \JJ \mid$. 
Let $\Delta \mu$ be the difference in chemical potential
between the two phases. 
By creating a droplet of size $L$ we both gain,
($\phi\Delta\mu(4/3)\pi L^3$) and loose
($-4\pi\sigma L^2$) free energy. 
Minimizing the change in free-energy we estimate
$\Delta \mu \approx 2\sigma/(L\phi) $. 
Typical gradients of chemical potential can then be estimated
as
$\grad\mu \sim \Delta\mu/L \sim 2\sigma/(\phi L^2)$
We assume that at all times the diffusive contribution to the flux, $J$,  in
\Eq{eq:flux} is negligible compared to the advective contribution.
In addition, at short times the flow
velocities are so small that \Eq{eq:CH} reduces to,
\begin{equation}
\frac{dL}{dt} \approx 0\quad {\rm and}\quad 
\partial_t \vv \approx \phi\grad\mu \sim 2\frac{\sigma}{L^2}\/.
\label{eq:shortt}
\end{equation}
Then at short times $L(t)$ remains almost a constant
and $E(t) \sim t^2$ -- this rationalizes the short-time behaviour
in \subfig{fig:growth}{B}.

At intermediate times and at large scales the contribution from
the nonlinear term dominates over the viscous term in the
momentum equation, \Eq{eq:mom}, such that 
\begin{equation}
  \frac{dL}{dt} \approx \phi v \quad {\rm and}\quad
  \rho\vv\cdot\grad\vv \sim \phi\grad\mu
  \label{eq:intert}
\end{equation}
Next we assume that $\ell$ is the characteristic
length scale of the flow.
If $\ell$ is proportional to  the scale of droplet, $L$,
$ v \sim \sqrt{\sigma/(\rho L)}$; consequently 
$ \sqrt{L}dL/dt \sim \sqrt{\sigma/\rho}$,
which in turn implies $L \sim t^{2/3}$ -- 
Furukawa's scaling~\cite{furukawa1985effect}.
Crucially, if $\ell$ is different from $L$ and
is almost a constant at intermediate times we obtain
$ L dL/dt \sim \sqrt{2\sigma\ell/\rho}$,
which implies $L \sim \sqrt{t}$.

\begin{figure}
  \includegraphics[width=0.95\columnwidth]{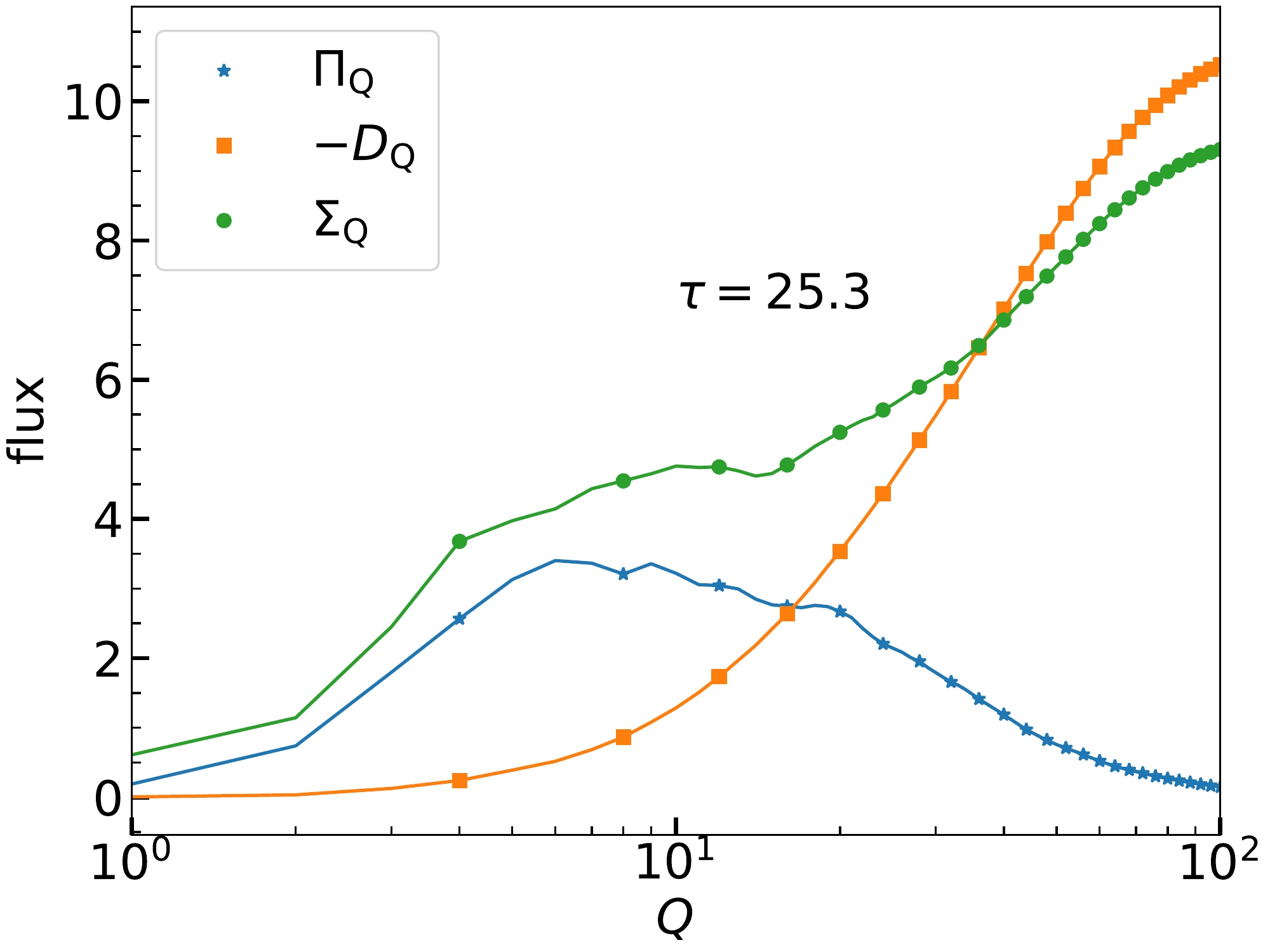}
  \caption{\label{fig:flux} The three terms on the right hand side of the
    energy flux equation~\eq{eq:eflux} at an intermediate time $\tau=25.3$.
    }
\end{figure}

Let us now critically examine our theory.
First, we assume the advective flux dominates over the diffusive
flux in \Eq{eq:flux}.
We find  that this is true in our simulations at all
times~\footnote{see Appendix~\ref{sec:flux}}. 

Second, at intermediate times ($10 \lesssim \tau \lesssim 100$)
and at large length scales the nonlinear term in \Eq{eq:mom}:
(a) dominates over the viscous term;
(b) has a characteristic length scale, $\ell$, different from
the scale $L$; and
(c) the length scale $\ell$ is almost a constant over the timescale.
If (a) holds then at large scales the flow must be  turbulent and
obeys Kolmogorov theory~\cite{Fri96}, i.e., 
there must exist a range of wavevectors 
$q$ over which $\bE(q) \sim q^{-5/3}$.
This is indeed what we have already shown in \Fig{fig:specu}.
Furthermore we notice that the spectra at large scale has
a characteristic peak near $q = \qstar \approx 9$
(marked by a vertical line in the inset of \Fig{fig:specu}).
The location of this peak, $\qstar$ does not change with time. 
Hence we can define the characteristic length scale
of the flow to be $\ell = 2\pi/q_{\ast}$ -- this scale remains
almost a constant during the intermediate times.
This confirms both (b) and (c) above. 

To directly examine the relative importance of the terms in the
momentum equation~\Eq{eq:mom}
it is usual~\cite{Fri96} to examine the scale-by-scale energy budget in the
Fourier space, defined by
\begin{equation}
  \EQ = {1\over 2}\int_0^Q dq\int \bE(q) \quad\/.
\label{eq:EQ}
\end{equation}
Straightforward algebra starting from \Eq{eq:mom} shows
\begin{equation}
    \partial_t \EQ = \PiQ + \DQ - \SigmaQ
\label{eq:eflux}
\end{equation}
where $\PiQ$, $\DQ$ and $\SigmaQ$ are contributions from the
nonlinear term, the viscous term,
and the $\phi\grad\mu$ in \Eq{eq:mom}, respectively~\footnote{ 
Due to the assumption of incompressibility the pressure gradient does
not make any contribution.}.
The viscous term $\DQ$ is always negative whereas
both $\PiQ$ and $\SigmaQ$ can, in principle, have either signs.
A positive $\PiQ$ implies that the kinetic energy cascades from
large to small length scales 
which is the case of Kolmogorov turbulence in three dimensions.
The contribution from surface tension $\SigmaQ$, is positive when
coagulation of droplets releases free energy that drives the
fluid and negative when the flow drives droplet motion. 
At short times, $\tau \lesssim 3$, as expected, we find that
both $\PiQ$ and $\DQ$ are small compared
to $\SigmaQ$~\footnote{see Appendix~\ref{sec:flux}}, hence
\eq{eq:eflux} reduces to $\delt \EQ \approx \SigmaQ$
-- supporting the dominant ballance assumed in \eq{eq:shortt}.
This implies $E(t) = \lim_{Q\to\infty} \mE_Q \sim t^2$ -- the same scaling
we find in \subfig{fig:growth}{B}. 
At intermediate times, for large lengh scale (small enough $Q$,  $Q<10$)
the contribution from $\phi\grad\mu$ is largely ballanced by the
nonlinear term, see \Fig{fig:flux}.
Consequently, the dominant ballance in \eq{eq:eflux} is
$\SigmaQ \approx \PiQ$ justifying \eq{eq:intert}.

Note that turbulence in Cahn--Hilliard--Navier--Stokes equation
has been studied extensively but also exclusively
with an external stirring force generating and maintaining the
turbulence~\cite{perlekar2014spinodal,Onuki2002phase,ruiz1981turbulence,
  fan2016cascades, ray2011universality,pal2016binary}.
By contrast, in our case the turbulence is generated by coagulating droplets.
The spectra for both velocity and the phase-variable, $\phi$, shows
power-law scaling, i.e., they are scale invariant.
Hence each possess only two characteristic length scales, respectively,
the large scale and small scale cutoffs of the scaling
which are often called the integral scale and the dissipative 
scale~\footnote{Here we ignore the multiscale/multifractal 
nature of turbulent fluctuations
which, to the best of our knowledge, has never been studied
in Cahn--Hilliard--Navier--Stokes equations.} . 
We use $\Sc = 1$, consequently the small scale cutoffs are
practically the same.
But the large scale cutoffs  are very different.

Two related problem are \textit{pseudo-turbulence} --
turbulence generated by rising
bubbles~\cite{lance1991turbulence,mudde2005gravity,prakash2016energy,
  risso2018agitation} in a quiescent fluid -- and
stirred Kolmogorov turbulence modified by rising
bubbles~\cite{lance1991turbulence,mathai2020bubble,pandey2020liquid,
  pandey2021turbulence}.
In both of these cases, coagulation plays negligible role and the
effect of the bubbles give rise to the kinetic energy spectrum
$\bE(q) \sim q^{-3}$~\footnote{
  In bubble--generated--turbulence,
  typically~\cite{lance1991turbulence,pandey2020liquid}
a $k^{-3}$ spectrum is observed.
In bubble--modified--turbulence, at small wavenumbers a $k^{-5/3}$
scaling is found which at large wavenumbers crosses over to $k^{-3}$
scaling~\cite{pandey2021turbulence}.}
that can be understood by balancing the energy production by the bubbles
with the viscous dissipation~\cite{lance1991turbulence,
  prakash2016energy,pandey2020liquid,pandey2021turbulence}.
Our simulations corresponds to a very different parameter range
where the contribution due to the chemical potential balances the
advective nonlinearity.

It has been emphasized before~\cite{wagner1998breakdown} that
coarsening of the domains in phase-separating binary fluids
is not scale-invariant -- different characteristic length scales
constructed from $\phi$ scales differently.
Note that we focus on a very different asepct -- we consider only the
integral scale, defined in \eq{eq:kandl},
and show that it scales as $L \sim t^{1/2}$ at intermediate times
during which the characteristic length scale of the flow remains
almost the same. 

In conclusion, we explore the coarsening phenomena at small
diffusivity and viscosity in that part of the phase diagram
where droplet spinodal decomposition operates.
We find turbulence a-la Kolmogorov and
emergence of very different characteristic scales of
the droplets and the flow -- general scale invariance
is broken -- the integral scale of the droplets
$L \sim t^{1/2}$ where the characteristic scale
of the flow remains practically constant. 
Nevertheless, we are able to generalize the standard
scaling theory to the present case and obtain
clear agreement between theory and simulations. 
The key to understand the problem is to calculate the
scale-by-scale energy budget.

\acknowledgements
The figures in this paper are plotted using the free software
matplotlib~\citep{Hunter:2007}
and VisIt~\citep{HPV:VisIt}.  
This work is partially  funded  by  the  ``Bottlenecks  for particle
growth  in  turbulent aerosols''  grant  from  the  Knut  and  Alice
Wallenberg  Foundation  (2014.0048).
PP  acknowledges support
from the Department of Atomic Energy (DAE), India under Project
Identification No. RTI 4007, and DST (India) Project
Nos. ECR/2018/001135 and DST/NSM/R\&D\_HPC\_Applications/2021/29.
DM acknowledges the support of the Swedish Research Council
Grant No. 638-2013-9243 as well as 2016-05225.
The simulations were performed on resources provided by
the Swedish National Infrastructure for Computing (SNIC) at PDC center for high performance
computing.
We particularly thank Tor Kjellsson and Harald Barth for their
help with visualization. 

\begin{thebibliography}{60}%
\makeatletter
\providecommand \@ifxundefined [1]{%
 \@ifx{#1\undefined}
}%
\providecommand \@ifnum [1]{%
 \ifnum #1\expandafter \@firstoftwo
 \else \expandafter \@secondoftwo
 \fi
}%
\providecommand \@ifx [1]{%
 \ifx #1\expandafter \@firstoftwo
 \else \expandafter \@secondoftwo
 \fi
}%
\providecommand \natexlab [1]{#1}%
\providecommand \enquote  [1]{``#1''}%
\providecommand \bibnamefont  [1]{#1}%
\providecommand \bibfnamefont [1]{#1}%
\providecommand \citenamefont [1]{#1}%
\providecommand \href@noop [0]{\@secondoftwo}%
\providecommand \href [0]{\begingroup \@sanitize@url \@href}%
\providecommand \@href[1]{\@@startlink{#1}\@@href}%
\providecommand \@@href[1]{\endgroup#1\@@endlink}%
\providecommand \@sanitize@url [0]{\catcode `\\12\catcode `\$12\catcode
  `\&12\catcode `\#12\catcode `\^12\catcode `\_12\catcode `\%12\relax}%
\providecommand \@@startlink[1]{}%
\providecommand \@@endlink[0]{}%
\providecommand \url  [0]{\begingroup\@sanitize@url \@url }%
\providecommand \@url [1]{\endgroup\@href {#1}{\urlprefix }}%
\providecommand \urlprefix  [0]{URL }%
\providecommand \Eprint [0]{\href }%
\providecommand \doibase [0]{http://dx.doi.org/}%
\providecommand \selectlanguage [0]{\@gobble}%
\providecommand \bibinfo  [0]{\@secondoftwo}%
\providecommand \bibfield  [0]{\@secondoftwo}%
\providecommand \translation [1]{[#1]}%
\providecommand \BibitemOpen [0]{}%
\providecommand \bibitemStop [0]{}%
\providecommand \bibitemNoStop [0]{.\EOS\space}%
\providecommand \EOS [0]{\spacefactor3000\relax}%
\providecommand \BibitemShut  [1]{\csname bibitem#1\endcsname}%
\let\auto@bib@innerbib\@empty
\bibitem [{\citenamefont {{Hohenberg}}\ and\ \citenamefont
  {{Halperin}}(1977)}]{hoh+hal77}%
  \BibitemOpen
  \bibfield  {author} {\bibinfo {author} {\bibfnamefont {P.~C.}\ \bibnamefont
  {{Hohenberg}}}\ and\ \bibinfo {author} {\bibfnamefont {B.~I.}\ \bibnamefont
  {{Halperin}}},\ }\bibfield  {title} {\enquote {\bibinfo {title} {Theory of
  dynamic critical phenomena},}\ }\href@noop {} {\bibfield  {journal} {\bibinfo
   {journal} {Rev. Mod. Phys.}\ }\textbf {\bibinfo {volume} {49}},\ \bibinfo
  {pages} {435} (\bibinfo {year} {1977})}\BibitemShut {NoStop}%
\bibitem [{\citenamefont {Chaikin}\ and\ \citenamefont
  {Lubensky}(1998)}]{Cha+Lub98}%
  \BibitemOpen
  \bibfield  {author} {\bibinfo {author} {\bibfnamefont {P.M.}\ \bibnamefont
  {Chaikin}}\ and\ \bibinfo {author} {\bibfnamefont {T.C.}\ \bibnamefont
  {Lubensky}},\ }\href@noop {} \textit {{\bibinfo {title} {Principles of
  condensed matter physics}}}\ (\bibinfo  {publisher} {Cambridge},\ \bibinfo
  {address} {Cambridge University Press, UK},\ \bibinfo {year}
  {1998})\BibitemShut {NoStop}%
\bibitem [{\citenamefont {Cates}\ and\ \citenamefont
  {Evans}(2000)}]{cates2000soft}%
  \BibitemOpen
  \bibfield  {author} {\bibinfo {author} {\bibfnamefont {Michael~E}\
  \bibnamefont {Cates}}\ and\ \bibinfo {author} {\bibfnamefont {Martin~R}\
  \bibnamefont {Evans}},\ }\href@noop {} \textit{ {\bibinfo {title} {Soft and
  fragile matter: nonequilibrium dynamics, metastability and flow (PBK)}}}\
  (\bibinfo  {publisher} {CRC Press},\ \bibinfo {year} {2000})\BibitemShut
  {NoStop}%
\bibitem [{\citenamefont {Bray}(2002)}]{bray2002theory}%
  \BibitemOpen
  \bibfield  {author} {\bibinfo {author} {\bibfnamefont {Alan~J}\ \bibnamefont
  {Bray}},\ }\bibfield  {title} {\enquote {\bibinfo {title} {Theory of
  phase-ordering kinetics},}\ }\href@noop {} {\bibfield  {journal} {\bibinfo
  {journal} {Advances in Physics}\ }\textbf {\bibinfo {volume} {51}},\ \bibinfo
  {pages} {481--587} (\bibinfo {year} {2002})}\BibitemShut {NoStop}%
\bibitem [{\citenamefont {Onuki}(2002)}]{Onuki2002phase}%
  \BibitemOpen
  \bibfield  {author} {\bibinfo {author} {\bibfnamefont {Akira}\ \bibnamefont
  {Onuki}},\ }\href@noop {} \textit{ {\bibinfo {title} {Phase transition
  dynamics}}}\ (\bibinfo  {publisher} {Cambridge University Press},\ \bibinfo
  {year} {2002})\BibitemShut {NoStop}%
\bibitem [{\citenamefont {Cates}(2017)}]{cates2017complex}%
  \BibitemOpen
  \bibfield  {author} {\bibinfo {author} {\bibfnamefont {ME}~\bibnamefont
  {Cates}},\ }\bibfield  {title} {\enquote {\bibinfo {title} {Complex fluids:
  the physics of emulsions},}\ }\href@noop {} {\bibfield  {journal} {\bibinfo
  {journal} {Soft Interfaces: Lecture Notes of the Les Houches Summer School:
  Volume 98, July 2012}\ }\textbf {\bibinfo {volume} {98}},\ \bibinfo {pages}
  {317} (\bibinfo {year} {2017})}\BibitemShut {NoStop}%
\bibitem [{\citenamefont {Cates}\ and\ \citenamefont
  {Tjhung}(2018)}]{cates2018theories}%
  \BibitemOpen
  \bibfield  {author} {\bibinfo {author} {\bibfnamefont {Michael~E}\
  \bibnamefont {Cates}}\ and\ \bibinfo {author} {\bibfnamefont {Elsen}\
  \bibnamefont {Tjhung}},\ }\bibfield  {title} {\enquote {\bibinfo {title}
  {Theories of binary fluid mixtures: from phase-separation kinetics to active
  emulsions},}\ }\href@noop {} {\bibfield  {journal} {\bibinfo  {journal}
  {Journal of Fluid Mechanics}\ }\textbf {\bibinfo {volume} {836}} (\bibinfo
  {year} {2018})}\BibitemShut {NoStop}%
\bibitem [{\citenamefont {Hoffer}\ and\ \citenamefont
  {Sinha}(1986)}]{hoffer1986dynamics}%
  \BibitemOpen
  \bibfield  {author} {\bibinfo {author} {\bibfnamefont {James~K}\ \bibnamefont
  {Hoffer}}\ and\ \bibinfo {author} {\bibfnamefont {Dipen~N}\ \bibnamefont
  {Sinha}},\ }\bibfield  {title} {\enquote {\bibinfo {title} {Dynamics of
  binary phase separation in liquid $^3$he--$^4$he mixtures},}\ }\href@noop {}
  {\bibfield  {journal} {\bibinfo  {journal} {Physical Review A}\ }\textbf
  {\bibinfo {volume} {33}},\ \bibinfo {pages} {1918} (\bibinfo {year}
  {1986})}\BibitemShut {NoStop}%
\bibitem [{\citenamefont {Hofmann}\ \emph {et~al.}(2014)\citenamefont
  {Hofmann}, \citenamefont {Natu},\ and\ \citenamefont
  {Sarma}}]{hofmann2014coarsening}%
  \BibitemOpen
  \bibfield  {author} {\bibinfo {author} {\bibfnamefont {Johannes}\
  \bibnamefont {Hofmann}}, \bibinfo {author} {\bibfnamefont {Stefan~S}\
  \bibnamefont {Natu}}, \ and\ \bibinfo {author} {\bibfnamefont {S~Das}\
  \bibnamefont {Sarma}},\ }\bibfield  {title} {\enquote {\bibinfo {title}
  {Coarsening dynamics of binary bose condensates},}\ }\href@noop {} {\bibfield
   {journal} {\bibinfo  {journal} {Physical review letters}\ }\textbf {\bibinfo
  {volume} {113}},\ \bibinfo {pages} {095702} (\bibinfo {year}
  {2014})}\BibitemShut {NoStop}%
\bibitem [{\citenamefont {Mendon{\c{c}}a}\ and\ \citenamefont
  {Kaiser}(2012)}]{mendoncca2012photon}%
  \BibitemOpen
  \bibfield  {author} {\bibinfo {author} {\bibfnamefont {JT}~\bibnamefont
  {Mendon{\c{c}}a}}\ and\ \bibinfo {author} {\bibfnamefont {R}~\bibnamefont
  {Kaiser}},\ }\bibfield  {title} {\enquote {\bibinfo {title} {Photon bubbles
  in ultracold matter},}\ }\href@noop {} {\bibfield  {journal} {\bibinfo
  {journal} {Physical review letters}\ }\textbf {\bibinfo {volume} {108}},\
  \bibinfo {pages} {033001} (\bibinfo {year} {2012})}\BibitemShut {NoStop}%
\bibitem [{\citenamefont {Chomaz}\ \emph {et~al.}(2004)\citenamefont {Chomaz},
  \citenamefont {Colonna},\ and\ \citenamefont {Randrup}}]{chomaz2004nuclear}%
  \BibitemOpen
  \bibfield  {author} {\bibinfo {author} {\bibfnamefont {Philippe}\
  \bibnamefont {Chomaz}}, \bibinfo {author} {\bibfnamefont {Maria}\
  \bibnamefont {Colonna}}, \ and\ \bibinfo {author} {\bibfnamefont {J{\o}rgen}\
  \bibnamefont {Randrup}},\ }\bibfield  {title} {\enquote {\bibinfo {title}
  {Nuclear spinodal fragmentation},}\ }\href@noop {} {\bibfield  {journal}
  {\bibinfo  {journal} {Physics reports}\ }\textbf {\bibinfo {volume} {389}},\
  \bibinfo {pages} {263--440} (\bibinfo {year} {2004})}\BibitemShut {NoStop}%
\bibitem [{\citenamefont {Boyanovsky}\ \emph {et~al.}(2006)\citenamefont
  {Boyanovsky}, \citenamefont {De~Vega},\ and\ \citenamefont
  {Schwarz}}]{boyanovsky2006phase}%
  \BibitemOpen
  \bibfield  {author} {\bibinfo {author} {\bibfnamefont {D}~\bibnamefont
  {Boyanovsky}}, \bibinfo {author} {\bibfnamefont {HJ}~\bibnamefont {De~Vega}},
  \ and\ \bibinfo {author} {\bibfnamefont {DJ}~\bibnamefont {Schwarz}},\
  }\bibfield  {title} {\enquote {\bibinfo {title} {Phase transitions in the
  early and present universe},}\ }\href@noop {} {\bibfield  {journal} {\bibinfo
   {journal} {Annu. Rev. Nucl. Part. Sci.}\ }\textbf {\bibinfo {volume} {56}},\
  \bibinfo {pages} {441--500} (\bibinfo {year} {2006})}\BibitemShut {NoStop}%
\bibitem [{\citenamefont {Prendergast}\ and\ \citenamefont
  {Spiegel}(1973)}]{prendergast1973photon}%
  \BibitemOpen
  \bibfield  {author} {\bibinfo {author} {\bibfnamefont {KH}~\bibnamefont
  {Prendergast}}\ and\ \bibinfo {author} {\bibfnamefont {EA}~\bibnamefont
  {Spiegel}},\ }\bibfield  {title} {\enquote {\bibinfo {title} {Photon
  bubbles},}\ }\href@noop {} {\bibfield  {journal} {\bibinfo  {journal}
  {Comments on Astrophysics and Space Physics}\ }\textbf {\bibinfo {volume}
  {5}},\ \bibinfo {pages} {43} (\bibinfo {year} {1973})}\BibitemShut {NoStop}%
\bibitem [{\citenamefont {Rieutord}\ \emph {et~al.}(2006)\citenamefont
  {Rieutord}, \citenamefont {Dubrulle},\ and\ \citenamefont
  {Spiegel}}]{rieutord2006phenomenological}%
  \BibitemOpen
  \bibfield  {author} {\bibinfo {author} {\bibfnamefont {M}~\bibnamefont
  {Rieutord}}, \bibinfo {author} {\bibfnamefont {B}~\bibnamefont {Dubrulle}}, \
  and\ \bibinfo {author} {\bibfnamefont {EA}~\bibnamefont {Spiegel}},\
  }\bibfield  {title} {\enquote {\bibinfo {title} {Phenomenological
  photofluiddynamics},}\ }\href@noop {} {\bibfield  {journal} {\bibinfo
  {journal} {European Astronomical Society Publications Series}\ }\textbf
  {\bibinfo {volume} {21}},\ \bibinfo {pages} {127--145} (\bibinfo {year}
  {2006})}\BibitemShut {NoStop}%
\bibitem [{\citenamefont {Cahn}\ and\ \citenamefont
  {Hilliard}(1958)}]{cahn1958free}%
  \BibitemOpen
  \bibfield  {author} {\bibinfo {author} {\bibfnamefont {John~W}\ \bibnamefont
  {Cahn}}\ and\ \bibinfo {author} {\bibfnamefont {John~E}\ \bibnamefont
  {Hilliard}},\ }\bibfield  {title} {\enquote {\bibinfo {title} {Free energy of
  a nonuniform system. i. interfacial free energy},}\ }\href@noop {} {\bibfield
   {journal} {\bibinfo  {journal} {The Journal of chemical physics}\ }\textbf
  {\bibinfo {volume} {28}},\ \bibinfo {pages} {258--267} (\bibinfo {year}
  {1958})}\BibitemShut {NoStop}%
\bibitem [{\citenamefont {Leyvraz}(2003)}]{leyvraz2003scaling}%
  \BibitemOpen
  \bibfield  {author} {\bibinfo {author} {\bibfnamefont {Fran{\c{c}}ois}\
  \bibnamefont {Leyvraz}},\ }\bibfield  {title} {\enquote {\bibinfo {title}
  {Scaling theory and exactly solved models in the kinetics of irreversible
  aggregation},}\ }\href@noop {} {\bibfield  {journal} {\bibinfo  {journal}
  {Physics Reports}\ }\textbf {\bibinfo {volume} {383}},\ \bibinfo {pages}
  {95--212} (\bibinfo {year} {2003})}\BibitemShut {NoStop}%
\bibitem [{\citenamefont {Lifshitz}\ and\ \citenamefont
  {Slyozov}(1961)}]{lifshitz1961kinetics}%
  \BibitemOpen
  \bibfield  {author} {\bibinfo {author} {\bibfnamefont {Ilya~M}\ \bibnamefont
  {Lifshitz}}\ and\ \bibinfo {author} {\bibfnamefont {Vitaly~V}\ \bibnamefont
  {Slyozov}},\ }\bibfield  {title} {\enquote {\bibinfo {title} {The kinetics of
  precipitation from supersaturated solid solutions},}\ }\href@noop {}
  {\bibfield  {journal} {\bibinfo  {journal} {Journal of physics and chemistry
  of solids}\ }\textbf {\bibinfo {volume} {19}},\ \bibinfo {pages} {35--50}
  (\bibinfo {year} {1961})}\BibitemShut {NoStop}%
\bibitem [{\citenamefont {Wagner}(1961)}]{wagner1961theory}%
  \BibitemOpen
  \bibfield  {author} {\bibinfo {author} {\bibfnamefont {C}~\bibnamefont
  {Wagner}},\ }\bibfield  {title} {\enquote {\bibinfo {title} {Theory of the
  aging of precipitates by dissolution-reprecipitation (ostwald ripening)},}\
  }\href@noop {} {\bibfield  {journal} {\bibinfo  {journal} {Z Elektrochem}\
  }\textbf {\bibinfo {volume} {65}},\ \bibinfo {pages} {581--11} (\bibinfo
  {year} {1961})}\BibitemShut {NoStop}%
\bibitem [{\citenamefont {Binder}\ and\ \citenamefont
  {Stauffer}(1974)}]{binder1974theory}%
  \BibitemOpen
  \bibfield  {author} {\bibinfo {author} {\bibfnamefont {Kurt}\ \bibnamefont
  {Binder}}\ and\ \bibinfo {author} {\bibfnamefont {D}~\bibnamefont
  {Stauffer}},\ }\bibfield  {title} {\enquote {\bibinfo {title} {Theory for the
  slowing down of the relaxation and spinodal decomposition of binary
  mixtures},}\ }\href@noop {} {\bibfield  {journal} {\bibinfo  {journal}
  {Physical Review Letters}\ }\textbf {\bibinfo {volume} {33}},\ \bibinfo
  {pages} {1006} (\bibinfo {year} {1974})}\BibitemShut {NoStop}%
\bibitem [{\citenamefont {Siggia}(1979)}]{siggia1979late}%
  \BibitemOpen
  \bibfield  {author} {\bibinfo {author} {\bibfnamefont {Eric~D}\ \bibnamefont
  {Siggia}},\ }\bibfield  {title} {\enquote {\bibinfo {title} {Late stages of
  spinodal decomposition in binary mixtures},}\ }\href@noop {} {\bibfield
  {journal} {\bibinfo  {journal} {Physical review A}\ }\textbf {\bibinfo
  {volume} {20}},\ \bibinfo {pages} {595} (\bibinfo {year} {1979})}\BibitemShut
  {NoStop}%
\bibitem [{\citenamefont {San~Miguel}\ \emph {et~al.}(1985)\citenamefont
  {San~Miguel}, \citenamefont {Grant},\ and\ \citenamefont
  {Gunton}}]{san1985phase}%
  \BibitemOpen
  \bibfield  {author} {\bibinfo {author} {\bibfnamefont {Maxi}\ \bibnamefont
  {San~Miguel}}, \bibinfo {author} {\bibfnamefont {Martin}\ \bibnamefont
  {Grant}}, \ and\ \bibinfo {author} {\bibfnamefont {James~D}\ \bibnamefont
  {Gunton}},\ }\bibfield  {title} {\enquote {\bibinfo {title} {Phase separation
  in two-dimensional binary fluids},}\ }\href@noop {} {\bibfield  {journal}
  {\bibinfo  {journal} {Physical Review A}\ }\textbf {\bibinfo {volume} {31}},\
  \bibinfo {pages} {1001} (\bibinfo {year} {1985})}\BibitemShut {NoStop}%
\bibitem [{\citenamefont {Tanaka}(1996)}]{tanaka1996coarsening}%
  \BibitemOpen
  \bibfield  {author} {\bibinfo {author} {\bibfnamefont {Hajime}\ \bibnamefont
  {Tanaka}},\ }\bibfield  {title} {\enquote {\bibinfo {title} {Coarsening
  mechanisms of droplet spinodal decomposition in binary fluid mixtures},}\
  }\href@noop {} {\bibfield  {journal} {\bibinfo  {journal} {The Journal of
  chemical physics}\ }\textbf {\bibinfo {volume} {105}},\ \bibinfo {pages}
  {10099--10114} (\bibinfo {year} {1996})}\BibitemShut {NoStop}%
\bibitem [{\citenamefont {Furukawa}(1985)}]{furukawa1985effect}%
  \BibitemOpen
  \bibfield  {author} {\bibinfo {author} {\bibfnamefont {Hiroshi}\ \bibnamefont
  {Furukawa}},\ }\bibfield  {title} {\enquote {\bibinfo {title} {Effect of
  inertia on droplet growth in a fluid},}\ }\href@noop {} {\bibfield  {journal}
  {\bibinfo  {journal} {Physical Review A}\ }\textbf {\bibinfo {volume} {31}},\
  \bibinfo {pages} {1103} (\bibinfo {year} {1985})}\BibitemShut {NoStop}%
\bibitem [{\citenamefont {Alexander}\ \emph {et~al.}(1993)\citenamefont
  {Alexander}, \citenamefont {Chen},\ and\ \citenamefont
  {Grunau}}]{alexander1993hydrodynamic}%
  \BibitemOpen
  \bibfield  {author} {\bibinfo {author} {\bibfnamefont {Francis~J}\
  \bibnamefont {Alexander}}, \bibinfo {author} {\bibfnamefont {Shiyi}\
  \bibnamefont {Chen}}, \ and\ \bibinfo {author} {\bibfnamefont {Daryl~W}\
  \bibnamefont {Grunau}},\ }\bibfield  {title} {\enquote {\bibinfo {title}
  {Hydrodynamic spinodal decomposition: Growth kinetics and scaling
  functions},}\ }\href@noop {} {\bibfield  {journal} {\bibinfo  {journal}
  {Physical Review B}\ }\textbf {\bibinfo {volume} {48}},\ \bibinfo {pages}
  {634} (\bibinfo {year} {1993})}\BibitemShut {NoStop}%
\bibitem [{\citenamefont {Kendon}\ \emph {et~al.}(2001)\citenamefont {Kendon},
  \citenamefont {Cates}, \citenamefont {Pagonabarraga}, \citenamefont
  {Desplat},\ and\ \citenamefont {Bladon}}]{kendon2001inertial}%
  \BibitemOpen
  \bibfield  {author} {\bibinfo {author} {\bibfnamefont {Vivien~M}\
  \bibnamefont {Kendon}}, \bibinfo {author} {\bibfnamefont {Michael~E}\
  \bibnamefont {Cates}}, \bibinfo {author} {\bibfnamefont {Ignacio}\
  \bibnamefont {Pagonabarraga}}, \bibinfo {author} {\bibfnamefont {J-C}\
  \bibnamefont {Desplat}}, \ and\ \bibinfo {author} {\bibfnamefont {Peter}\
  \bibnamefont {Bladon}},\ }\bibfield  {title} {\enquote {\bibinfo {title}
  {Inertial effects in three-dimensional spinodal decomposition of a symmetric
  binary fluid mixture: a lattice boltzmann study},}\ }\href@noop {} {\bibfield
   {journal} {\bibinfo  {journal} {Journal of Fluid Mechanics}\ }\textbf
  {\bibinfo {volume} {440}},\ \bibinfo {pages} {147--203} (\bibinfo {year}
  {2001})}\BibitemShut {NoStop}%
\bibitem [{\citenamefont {Farrell}\ and\ \citenamefont
  {Valls}(1991)}]{farrell1991growth}%
  \BibitemOpen
  \bibfield  {author} {\bibinfo {author} {\bibfnamefont {James~E}\ \bibnamefont
  {Farrell}}\ and\ \bibinfo {author} {\bibfnamefont {Oriol~T}\ \bibnamefont
  {Valls}},\ }\bibfield  {title} {\enquote {\bibinfo {title} {Growth kinetics
  and domain morphology after off-critical quenches in a two-dimensional fluid
  model},}\ }\href@noop {} {\bibfield  {journal} {\bibinfo  {journal} {Physical
  Review B}\ }\textbf {\bibinfo {volume} {43}},\ \bibinfo {pages} {630}
  (\bibinfo {year} {1991})}\BibitemShut {NoStop}%
\bibitem [{\citenamefont {Datt}\ \emph {et~al.}(2015)\citenamefont {Datt},
  \citenamefont {Thampi},\ and\ \citenamefont
  {Govindarajan}}]{datt2015morphological}%
  \BibitemOpen
  \bibfield  {author} {\bibinfo {author} {\bibfnamefont {Charu}\ \bibnamefont
  {Datt}}, \bibinfo {author} {\bibfnamefont {Sumesh~P}\ \bibnamefont {Thampi}},
  \ and\ \bibinfo {author} {\bibfnamefont {Rama}\ \bibnamefont
  {Govindarajan}},\ }\bibfield  {title} {\enquote {\bibinfo {title}
  {Morphological evolution of domains in spinodal decomposition},}\ }\href@noop
  {} {\bibfield  {journal} {\bibinfo  {journal} {Physical Review E}\ }\textbf
  {\bibinfo {volume} {91}},\ \bibinfo {pages} {010101} (\bibinfo {year}
  {2015})}\BibitemShut {NoStop}%
\bibitem [{\citenamefont {Shimizu}\ and\ \citenamefont
  {Tanaka}(2015)}]{shimizu2015novel}%
  \BibitemOpen
  \bibfield  {author} {\bibinfo {author} {\bibfnamefont {Ryotaro}\ \bibnamefont
  {Shimizu}}\ and\ \bibinfo {author} {\bibfnamefont {Hajime}\ \bibnamefont
  {Tanaka}},\ }\bibfield  {title} {\enquote {\bibinfo {title} {A novel
  coarsening mechanism of droplets in immiscible fluid mixtures},}\ }\href@noop
  {} {\bibfield  {journal} {\bibinfo  {journal} {Nature communications}\
  }\textbf {\bibinfo {volume} {6}},\ \bibinfo {pages} {1--11} (\bibinfo {year}
  {2015})}\BibitemShut {NoStop}%
\bibitem [{\citenamefont {Chou}\ and\ \citenamefont
  {Goldburg}(1979)}]{chou1979phase}%
  \BibitemOpen
  \bibfield  {author} {\bibinfo {author} {\bibfnamefont {Ya-Chang}\
  \bibnamefont {Chou}}\ and\ \bibinfo {author} {\bibfnamefont {Walter~I}\
  \bibnamefont {Goldburg}},\ }\href@noop {} {\enquote {\bibinfo {title} {Phase
  separation and coalescence in critically quenched isobutyric-acid—water and
  2, 6-lutidine—water mixtures},}\ } (\bibinfo {year} {1979})\BibitemShut
  {NoStop}%
\bibitem [{\citenamefont {Wong}\ and\ \citenamefont
  {Knobler}(1981)}]{wong1981light}%
  \BibitemOpen
  \bibfield  {author} {\bibinfo {author} {\bibfnamefont {Ning-Chih}\
  \bibnamefont {Wong}}\ and\ \bibinfo {author} {\bibfnamefont {Charles~M}\
  \bibnamefont {Knobler}},\ }\bibfield  {title} {\enquote {\bibinfo {title}
  {Light-scattering studies of phase separation in isobutyric acid+ water
  mixtures: Hydrodynamic effects},}\ }\href@noop {} {\bibfield  {journal}
  {\bibinfo  {journal} {Physical Review A}\ }\textbf {\bibinfo {volume} {24}},\
  \bibinfo {pages} {3205} (\bibinfo {year} {1981})}\BibitemShut {NoStop}%
\bibitem [{\citenamefont {Perrot}\ \emph {et~al.}(1994)\citenamefont {Perrot},
  \citenamefont {Guenoun}, \citenamefont {Baumberger}, \citenamefont {Beysens},
  \citenamefont {Garrabos},\ and\ \citenamefont
  {Le~Neindre}}]{perrot1994nucleation}%
  \BibitemOpen
  \bibfield  {author} {\bibinfo {author} {\bibfnamefont {F}~\bibnamefont
  {Perrot}}, \bibinfo {author} {\bibfnamefont {P}~\bibnamefont {Guenoun}},
  \bibinfo {author} {\bibfnamefont {T}~\bibnamefont {Baumberger}}, \bibinfo
  {author} {\bibfnamefont {D}~\bibnamefont {Beysens}}, \bibinfo {author}
  {\bibfnamefont {Y}~\bibnamefont {Garrabos}}, \ and\ \bibinfo {author}
  {\bibfnamefont {B}~\bibnamefont {Le~Neindre}},\ }\bibfield  {title} {\enquote
  {\bibinfo {title} {Nucleation and growth of tightly packed droplets in
  fluids},}\ }\href@noop {} {\bibfield  {journal} {\bibinfo  {journal}
  {Physical review letters}\ }\textbf {\bibinfo {volume} {73}},\ \bibinfo
  {pages} {688} (\bibinfo {year} {1994})}\BibitemShut {NoStop}%
\bibitem [{\citenamefont {Rahman}\ \emph {et~al.}(2019)\citenamefont {Rahman},
  \citenamefont {Lee}, \citenamefont {Iyer},\ and\ \citenamefont
  {Williams}}]{rahman2019viscous}%
  \BibitemOpen
  \bibfield  {author} {\bibinfo {author} {\bibfnamefont {Md~Mahmudur}\
  \bibnamefont {Rahman}}, \bibinfo {author} {\bibfnamefont {Willis}\
  \bibnamefont {Lee}}, \bibinfo {author} {\bibfnamefont {Arvind}\ \bibnamefont
  {Iyer}}, \ and\ \bibinfo {author} {\bibfnamefont {Stuart~J}\ \bibnamefont
  {Williams}},\ }\bibfield  {title} {\enquote {\bibinfo {title} {Viscous
  resistance in drop coalescence},}\ }\href@noop {} {\bibfield  {journal}
  {\bibinfo  {journal} {Physics of Fluids}\ }\textbf {\bibinfo {volume} {31}},\
  \bibinfo {pages} {012104} (\bibinfo {year} {2019})}\BibitemShut {NoStop}%
\bibitem [{\citenamefont {Naso}\ and\ \citenamefont
  {N{\'a}raigh}(2018)}]{naso2018flow}%
  \BibitemOpen
  \bibfield  {author} {\bibinfo {author} {\bibfnamefont {Aurore}\ \bibnamefont
  {Naso}}\ and\ \bibinfo {author} {\bibfnamefont {Lennon~{\'O}}\ \bibnamefont
  {N{\'a}raigh}},\ }\bibfield  {title} {\enquote {\bibinfo {title} {A
  flow-pattern map for phase separation using the
  navier--stokes--cahn--hilliard model},}\ }\href@noop {} {\bibfield  {journal}
  {\bibinfo  {journal} {European Journal of Mechanics-B/Fluids}\ }\textbf
  {\bibinfo {volume} {72}},\ \bibinfo {pages} {576--585} (\bibinfo {year}
  {2018})}\BibitemShut {NoStop}%
\bibitem [{Note1()}]{Note1}%
  \BibitemOpen
  \bibinfo {note} {A comprehensive description of the algorithm, the complete
  list of parameters and the non-dimensional equations are given in Appendix
  \ref {sec:MM}. A comparison of our parameters with Ref.~\cite {naso2018flow}
  and Ref.~\cite {kendon2001inertial} is given in Appendix \ref
  {sec:comparison}.}\BibitemShut {Stop}%
\bibitem [{\citenamefont {Perlekar}\ \emph {et~al.}(2014)\citenamefont
  {Perlekar}, \citenamefont {Benzi}, \citenamefont {Clercx}, \citenamefont
  {Nelson},\ and\ \citenamefont {Toschi}}]{perlekar2014spinodal}%
  \BibitemOpen
  \bibfield  {author} {\bibinfo {author} {\bibfnamefont {Prasad}\ \bibnamefont
  {Perlekar}}, \bibinfo {author} {\bibfnamefont {Roberto}\ \bibnamefont
  {Benzi}}, \bibinfo {author} {\bibfnamefont {Herman~JH}\ \bibnamefont
  {Clercx}}, \bibinfo {author} {\bibfnamefont {David~R}\ \bibnamefont
  {Nelson}}, \ and\ \bibinfo {author} {\bibfnamefont {Federico}\ \bibnamefont
  {Toschi}},\ }\bibfield  {title} {\enquote {\bibinfo {title} {Spinodal
  decomposition in homogeneous and isotropic turbulence},}\ }\href@noop {}
  {\bibfield  {journal} {\bibinfo  {journal} {Physical review letters}\
  }\textbf {\bibinfo {volume} {112}},\ \bibinfo {pages} {014502} (\bibinfo
  {year} {2014})}\BibitemShut {NoStop}%
\bibitem [{\citenamefont {Wu}\ \emph {et~al.}(1995)\citenamefont {Wu},
  \citenamefont {Alexander}, \citenamefont {Lookman},\ and\ \citenamefont
  {Chen}}]{wu1995effects}%
  \BibitemOpen
  \bibfield  {author} {\bibinfo {author} {\bibfnamefont {Yanan}\ \bibnamefont
  {Wu}}, \bibinfo {author} {\bibfnamefont {Francis~J}\ \bibnamefont
  {Alexander}}, \bibinfo {author} {\bibfnamefont {Turab}\ \bibnamefont
  {Lookman}}, \ and\ \bibinfo {author} {\bibfnamefont {Shiyi}\ \bibnamefont
  {Chen}},\ }\bibfield  {title} {\enquote {\bibinfo {title} {Effects of
  hydrodynamics on phase transition kinetics in two-dimensional binary
  fluids},}\ }\href@noop {} {\bibfield  {journal} {\bibinfo  {journal}
  {Physical review letters}\ }\textbf {\bibinfo {volume} {74}},\ \bibinfo
  {pages} {3852} (\bibinfo {year} {1995})}\BibitemShut {NoStop}%
\bibitem [{Note2()}]{Note2}%
  \BibitemOpen
  \bibinfo {note} {Ref~\cite {osborn1995lattice} has found the exponent $1/2$
  for liquid-gas systems, which is similar to model A of Hohenberg and
  Halperin, but not for binary fluids. Ref.~\cite {wagner1998breakdown} also
  obtained $1/2$ but for a length scale that is different from $L$ for a case
  with low viscosity where they found that general scale invariance is broken
  -- length scales defined in different ways scale differently. Both of these
  simulations are in two dimensions.}\BibitemShut {Stop}%
\bibitem [{\citenamefont {Frisch}(1996)}]{Fri96}%
  \BibitemOpen
  \bibfield  {author} {\bibinfo {author} {\bibfnamefont {U.}~\bibnamefont
  {Frisch}},\ }\href@noop {} \textit{ {\bibinfo {title} {Turbulence the legacy of
  A.N. Kolmogorov}}}\ (\bibinfo  {publisher} {Cambridge University Press},\
  \bibinfo {address} {Cambridge},\ \bibinfo {year} {1996})\BibitemShut
  {NoStop}%
\bibitem [{\citenamefont {Aurell}\ \emph {et~al.}(1992)\citenamefont {Aurell},
  \citenamefont {Frisch}, \citenamefont {Lutsko},\ and\ \citenamefont
  {Vergassola}}]{aurell1992multifractal}%
  \BibitemOpen
  \bibfield  {author} {\bibinfo {author} {\bibfnamefont {E}~\bibnamefont
  {Aurell}}, \bibinfo {author} {\bibfnamefont {Uriel}\ \bibnamefont {Frisch}},
  \bibinfo {author} {\bibfnamefont {James}\ \bibnamefont {Lutsko}}, \ and\
  \bibinfo {author} {\bibfnamefont {M}~\bibnamefont {Vergassola}},\ }\bibfield
  {title} {\enquote {\bibinfo {title} {On the multifractal properties of the
  energy dissipation derived from turbulence data},}\ }\href@noop {} {\bibfield
   {journal} {\bibinfo  {journal} {Journal of Fluid Mechanics}\ }\textbf
  {\bibinfo {volume} {238}},\ \bibinfo {pages} {467--486} (\bibinfo {year}
  {1992})}\BibitemShut {NoStop}%
\bibitem [{Note3()}]{Note3}%
  \BibitemOpen
  \bibinfo {note} {See Appendix~\ref {sec:flux}}\BibitemShut {NoStop}%
\bibitem [{Note4()}]{Note4}%
  \BibitemOpen
  \bibinfo {note} {Due to the assumption of incompressibility the pressure
  gradient does not make any contribution.}\BibitemShut {Stop}%
\bibitem [{Note5()}]{Note5}%
  \BibitemOpen
  \bibinfo {note} {See Appendix~\ref {sec:flux}}\BibitemShut {NoStop}%
\bibitem [{\citenamefont {Ruiz}\ and\ \citenamefont
  {Nelson}(1981)}]{ruiz1981turbulence}%
  \BibitemOpen
  \bibfield  {author} {\bibinfo {author} {\bibfnamefont {Ricardo}\ \bibnamefont
  {Ruiz}}\ and\ \bibinfo {author} {\bibfnamefont {David~R}\ \bibnamefont
  {Nelson}},\ }\bibfield  {title} {\enquote {\bibinfo {title} {Turbulence in
  binary fluid mixtures},}\ }\href@noop {} {\bibfield  {journal} {\bibinfo
  {journal} {Physical Review A}\ }\textbf {\bibinfo {volume} {23}},\ \bibinfo
  {pages} {3224} (\bibinfo {year} {1981})}\BibitemShut {NoStop}%
\bibitem [{\citenamefont {Fan}\ \emph {et~al.}(2016)\citenamefont {Fan},
  \citenamefont {Diamond}, \citenamefont {Chac{\'o}n},\ and\ \citenamefont
  {Li}}]{fan2016cascades}%
  \BibitemOpen
  \bibfield  {author} {\bibinfo {author} {\bibfnamefont {Xiang}\ \bibnamefont
  {Fan}}, \bibinfo {author} {\bibfnamefont {PH}~\bibnamefont {Diamond}},
  \bibinfo {author} {\bibfnamefont {L}~\bibnamefont {Chac{\'o}n}}, \ and\
  \bibinfo {author} {\bibfnamefont {Hui}\ \bibnamefont {Li}},\ }\bibfield
  {title} {\enquote {\bibinfo {title} {Cascades and spectra of a turbulent
  spinodal decomposition in two-dimensional symmetric binary liquid
  mixtures},}\ }\href@noop {} {\bibfield  {journal} {\bibinfo  {journal}
  {Physical Review Fluids}\ }\textbf {\bibinfo {volume} {1}},\ \bibinfo {pages}
  {054403} (\bibinfo {year} {2016})}\BibitemShut {NoStop}%
\bibitem [{\citenamefont {Ray}\ and\ \citenamefont
  {Basu}(2011)}]{ray2011universality}%
  \BibitemOpen
  \bibfield  {author} {\bibinfo {author} {\bibfnamefont {Samriddhi~Sankar}\
  \bibnamefont {Ray}}\ and\ \bibinfo {author} {\bibfnamefont {Abhik}\
  \bibnamefont {Basu}},\ }\bibfield  {title} {\enquote {\bibinfo {title}
  {Universality of scaling and multiscaling in turbulent symmetric binary
  fluids},}\ }\href@noop {} {\bibfield  {journal} {\bibinfo  {journal}
  {Physical Review E}\ }\textbf {\bibinfo {volume} {84}},\ \bibinfo {pages}
  {036316} (\bibinfo {year} {2011})}\BibitemShut {NoStop}%
\bibitem [{\citenamefont {Pal}\ \emph {et~al.}(2016)\citenamefont {Pal},
  \citenamefont {Perlekar}, \citenamefont {Gupta},\ and\ \citenamefont
  {Pandit}}]{pal2016binary}%
  \BibitemOpen
  \bibfield  {author} {\bibinfo {author} {\bibfnamefont {Nairita}\ \bibnamefont
  {Pal}}, \bibinfo {author} {\bibfnamefont {Prasad}\ \bibnamefont {Perlekar}},
  \bibinfo {author} {\bibfnamefont {Anupam}\ \bibnamefont {Gupta}}, \ and\
  \bibinfo {author} {\bibfnamefont {Rahul}\ \bibnamefont {Pandit}},\ }\bibfield
   {title} {\enquote {\bibinfo {title} {Binary-fluid turbulence: Signatures of
  multifractal droplet dynamics and dissipation reduction},}\ }\href@noop {}
  {\bibfield  {journal} {\bibinfo  {journal} {Physical Review E}\ }\textbf
  {\bibinfo {volume} {93}},\ \bibinfo {pages} {063115} (\bibinfo {year}
  {2016})}\BibitemShut {NoStop}%
\bibitem [{Note6()}]{Note6}%
  \BibitemOpen
  \bibinfo {note} {Here we ignore the multiscale/multifractal nature of
  turbulent fluctuations which, to the best of our knowledge, has never been
  studied in Cahn--Hilliard--Navier--Stokes equations.}\BibitemShut {Stop}%
\bibitem [{\citenamefont {Lance}\ and\ \citenamefont
  {Bataille}(1991)}]{lance1991turbulence}%
  \BibitemOpen
  \bibfield  {author} {\bibinfo {author} {\bibfnamefont {M}~\bibnamefont
  {Lance}}\ and\ \bibinfo {author} {\bibfnamefont {J}~\bibnamefont
  {Bataille}},\ }\bibfield  {title} {\enquote {\bibinfo {title} {Turbulence in
  the liquid phase of a uniform bubbly air--water flow},}\ }\href@noop {}
  {\bibfield  {journal} {\bibinfo  {journal} {Journal of fluid mechanics}\
  }\textbf {\bibinfo {volume} {222}},\ \bibinfo {pages} {95--118} (\bibinfo
  {year} {1991})}\BibitemShut {NoStop}%
\bibitem [{\citenamefont {Mudde}(2005)}]{mudde2005gravity}%
  \BibitemOpen
  \bibfield  {author} {\bibinfo {author} {\bibfnamefont {Robert~F}\
  \bibnamefont {Mudde}},\ }\bibfield  {title} {\enquote {\bibinfo {title}
  {Gravity-driven bubbly flows},}\ }\href@noop {} {\bibfield  {journal}
  {\bibinfo  {journal} {Annu. Rev. Fluid Mech.}\ }\textbf {\bibinfo {volume}
  {37}},\ \bibinfo {pages} {393--423} (\bibinfo {year} {2005})}\BibitemShut
  {NoStop}%
\bibitem [{\citenamefont {Prakash}\ \emph {et~al.}(2016)\citenamefont
  {Prakash}, \citenamefont {Mercado}, \citenamefont {van Wijngaarden},
  \citenamefont {Mancilla}, \citenamefont {Tagawa}, \citenamefont {Lohse},\
  and\ \citenamefont {Sun}}]{prakash2016energy}%
  \BibitemOpen
  \bibfield  {author} {\bibinfo {author} {\bibfnamefont {Vivek~N}\ \bibnamefont
  {Prakash}}, \bibinfo {author} {\bibfnamefont {J~Mart{\'\i}nez}\ \bibnamefont
  {Mercado}}, \bibinfo {author} {\bibfnamefont {Leen}\ \bibnamefont {van
  Wijngaarden}}, \bibinfo {author} {\bibfnamefont {Ernesto}\ \bibnamefont
  {Mancilla}}, \bibinfo {author} {\bibfnamefont {Yoshiyuki}\ \bibnamefont
  {Tagawa}}, \bibinfo {author} {\bibfnamefont {Detlef}\ \bibnamefont {Lohse}},
  \ and\ \bibinfo {author} {\bibfnamefont {Chao}\ \bibnamefont {Sun}},\
  }\bibfield  {title} {\enquote {\bibinfo {title} {Energy spectra in turbulent
  bubbly flows},}\ }\href@noop {} {\bibfield  {journal} {\bibinfo  {journal}
  {Journal of fluid mechanics}\ }\textbf {\bibinfo {volume} {791}},\ \bibinfo
  {pages} {174--190} (\bibinfo {year} {2016})}\BibitemShut {NoStop}%
\bibitem [{\citenamefont {Risso}(2018)}]{risso2018agitation}%
  \BibitemOpen
  \bibfield  {author} {\bibinfo {author} {\bibfnamefont {Fr{\'e}d{\'e}ric}\
  \bibnamefont {Risso}},\ }\bibfield  {title} {\enquote {\bibinfo {title}
  {Agitation, mixing, and transfers induced by bubbles},}\ }\href@noop {}
  {\bibfield  {journal} {\bibinfo  {journal} {Annual Review of Fluid
  Mechanics}\ }\textbf {\bibinfo {volume} {50}},\ \bibinfo {pages} {25--48}
  (\bibinfo {year} {2018})}\BibitemShut {NoStop}%
\bibitem [{\citenamefont {Mathai}\ \emph {et~al.}(2020)\citenamefont {Mathai},
  \citenamefont {Lohse},\ and\ \citenamefont {Sun}}]{mathai2020bubble}%
  \BibitemOpen
  \bibfield  {author} {\bibinfo {author} {\bibfnamefont {Varghese}\
  \bibnamefont {Mathai}}, \bibinfo {author} {\bibfnamefont {Detlef}\
  \bibnamefont {Lohse}}, \ and\ \bibinfo {author} {\bibfnamefont {Chao}\
  \bibnamefont {Sun}},\ }\bibfield  {title} {\enquote {\bibinfo {title} {Bubble
  and buoyant particle laden turbulent flows},}\ }\href@noop {} {\bibfield
  {journal} {\bibinfo  {journal} {Annu. Rev. Condens. Matter Phys}\ }\textbf
  {\bibinfo {volume} {11}} (\bibinfo {year} {2020})}\BibitemShut {NoStop}%
\bibitem [{\citenamefont {Pandey}\ \emph {et~al.}(2020)\citenamefont {Pandey},
  \citenamefont {Ramadugu},\ and\ \citenamefont {Perlekar}}]{pandey2020liquid}%
  \BibitemOpen
  \bibfield  {author} {\bibinfo {author} {\bibfnamefont {Vikash}\ \bibnamefont
  {Pandey}}, \bibinfo {author} {\bibfnamefont {Rashmi}\ \bibnamefont
  {Ramadugu}}, \ and\ \bibinfo {author} {\bibfnamefont {Prasad}\ \bibnamefont
  {Perlekar}},\ }\bibfield  {title} {\enquote {\bibinfo {title} {Liquid
  velocity fluctuations and energy spectra in three-dimensional buoyancy-driven
  bubbly flows},}\ }\href@noop {} {\bibfield  {journal} {\bibinfo  {journal}
  {Journal of Fluid Mechanics}\ }\textbf {\bibinfo {volume} {884}} (\bibinfo
  {year} {2020})}\BibitemShut {NoStop}%
\bibitem [{\citenamefont {Pandey}\ \emph {et~al.}(2021)\citenamefont {Pandey},
  \citenamefont {Mitra},\ and\ \citenamefont
  {Perlekar}}]{pandey2021turbulence}%
  \BibitemOpen
  \bibfield  {author} {\bibinfo {author} {\bibfnamefont {Vikash}\ \bibnamefont
  {Pandey}}, \bibinfo {author} {\bibfnamefont {Dhrubaditya}\ \bibnamefont
  {Mitra}}, \ and\ \bibinfo {author} {\bibfnamefont {Prasad}\ \bibnamefont
  {Perlekar}},\ }\bibfield  {title} {\enquote {\bibinfo {title} {Turbulence
  modulation in buoyancy-driven bubbly flows},}\ }\href@noop {} {\bibfield
  {journal} {\bibinfo  {journal} {arXiv preprint arXiv:2105.04812}\ } (\bibinfo
  {year} {2021})}\BibitemShut {NoStop}%
\bibitem [{Note7()}]{Note7}%
  \BibitemOpen
  \bibinfo {note} {In bubble--generated--turbulence, typically~\cite
  {lance1991turbulence,pandey2020liquid} a $k^{-3}$ spectrum is observed. In
  bubble--modified--turbulence, at small wavenumbers a $k^{-5/3}$ scaling is
  found which at large wavenumbers crosses over to $k^{-3}$ scaling~\cite
  {pandey2021turbulence}.}\BibitemShut {Stop}%
\bibitem [{\citenamefont {Wagner}\ and\ \citenamefont
  {Yeomans}(1998)}]{wagner1998breakdown}%
  \BibitemOpen
  \bibfield  {author} {\bibinfo {author} {\bibfnamefont {Alexander~J}\
  \bibnamefont {Wagner}}\ and\ \bibinfo {author} {\bibfnamefont
  {JM}~\bibnamefont {Yeomans}},\ }\bibfield  {title} {\enquote {\bibinfo
  {title} {Breakdown of scale invariance in the coarsening of phase-separating
  binary fluids},}\ }\href@noop {} {\bibfield  {journal} {\bibinfo  {journal}
  {Physical Review Letters}\ }\textbf {\bibinfo {volume} {80}},\ \bibinfo
  {pages} {1429} (\bibinfo {year} {1998})}\BibitemShut {NoStop}%
\bibitem [{\citenamefont {Hunter}(2007)}]{Hunter:2007}%
  \BibitemOpen
  \bibfield  {author} {\bibinfo {author} {\bibfnamefont {J.~D.}\ \bibnamefont
  {Hunter}},\ }\bibfield  {title} {\enquote {\bibinfo {title} {Matplotlib: A 2d
  graphics environment},}\ }\href {\doibase 10.1109/MCSE.2007.55} {\bibfield
  {journal} {\bibinfo  {journal} {Computing in Science \& Engineering}\
  }\textbf {\bibinfo {volume} {9}},\ \bibinfo {pages} {90--95} (\bibinfo {year}
  {2007})}\BibitemShut {NoStop}%
\bibitem [{\citenamefont {Childs}\ \emph {et~al.}(2012)\citenamefont {Childs},
  \citenamefont {Brugger}, \citenamefont {Whitlock}, \citenamefont {Meredith},
  \citenamefont {Ahern}, \citenamefont {Pugmire}, \citenamefont {Biagas},
  \citenamefont {Miller}, \citenamefont {Harrison}, \citenamefont {Weber},
  \citenamefont {Krishnan}, \citenamefont {Fogal}, \citenamefont {Sanderson},
  \citenamefont {Garth}, \citenamefont {Bethel}, \citenamefont {Camp},
  \citenamefont {R\"{u}bel}, \citenamefont {Durant}, \citenamefont {Favre},\
  and\ \citenamefont {Navr\'{a}til}}]{HPV:VisIt}%
  \BibitemOpen
  \bibfield  {author} {\bibinfo {author} {\bibfnamefont {Hank}\ \bibnamefont
  {Childs}}, \bibinfo {author} {\bibfnamefont {Eric}\ \bibnamefont {Brugger}},
  \bibinfo {author} {\bibfnamefont {Brad}\ \bibnamefont {Whitlock}}, \bibinfo
  {author} {\bibfnamefont {Jeremy}\ \bibnamefont {Meredith}}, \bibinfo {author}
  {\bibfnamefont {Sean}\ \bibnamefont {Ahern}}, \bibinfo {author}
  {\bibfnamefont {David}\ \bibnamefont {Pugmire}}, \bibinfo {author}
  {\bibfnamefont {Kathleen}\ \bibnamefont {Biagas}}, \bibinfo {author}
  {\bibfnamefont {Mark}\ \bibnamefont {Miller}}, \bibinfo {author}
  {\bibfnamefont {Cyrus}\ \bibnamefont {Harrison}}, \bibinfo {author}
  {\bibfnamefont {Gunther~H.}\ \bibnamefont {Weber}}, \bibinfo {author}
  {\bibfnamefont {Hari}\ \bibnamefont {Krishnan}}, \bibinfo {author}
  {\bibfnamefont {Thomas}\ \bibnamefont {Fogal}}, \bibinfo {author}
  {\bibfnamefont {Allen}\ \bibnamefont {Sanderson}}, \bibinfo {author}
  {\bibfnamefont {Christoph}\ \bibnamefont {Garth}}, \bibinfo {author}
  {\bibfnamefont {E.~Wes}\ \bibnamefont {Bethel}}, \bibinfo {author}
  {\bibfnamefont {David}\ \bibnamefont {Camp}}, \bibinfo {author}
  {\bibfnamefont {Oliver}\ \bibnamefont {R\"{u}bel}}, \bibinfo {author}
  {\bibfnamefont {Marc}\ \bibnamefont {Durant}}, \bibinfo {author}
  {\bibfnamefont {Jean~M.}\ \bibnamefont {Favre}}, \ and\ \bibinfo {author}
  {\bibfnamefont {Paul}\ \bibnamefont {Navr\'{a}til}},\ }\bibfield  {title}
  {\enquote {\bibinfo {title} {{VisIt: An End-User Tool For Visualizing and
  Analyzing Very Large Data}},}\ }in\ \href@noop {} \textit{ {\bibinfo
  {booktitle} {{High Performance Visualization--Enabling Extreme-Scale
  Scientific Insight}}}}\ (\bibinfo {year} {2012})\ pp.\ \bibinfo {pages}
  {357--372}\BibitemShut {NoStop}%
\bibitem [{\citenamefont {Osborn}\ \emph {et~al.}(1995)\citenamefont {Osborn},
  \citenamefont {Orlandini}, \citenamefont {Swift}, \citenamefont {Yeomans},\
  and\ \citenamefont {Banavar}}]{osborn1995lattice}%
  \BibitemOpen
  \bibfield  {author} {\bibinfo {author} {\bibfnamefont {WR}~\bibnamefont
  {Osborn}}, \bibinfo {author} {\bibfnamefont {E}~\bibnamefont {Orlandini}},
  \bibinfo {author} {\bibfnamefont {Michael~R}\ \bibnamefont {Swift}}, \bibinfo
  {author} {\bibfnamefont {JM}~\bibnamefont {Yeomans}}, \ and\ \bibinfo
  {author} {\bibfnamefont {Jayanth~R}\ \bibnamefont {Banavar}},\ }\bibfield
  {title} {\enquote {\bibinfo {title} {Lattice boltzmann study of hydrodynamic
  spinodal decomposition},}\ }\href@noop {} {\bibfield  {journal} {\bibinfo
  {journal} {Physical review letters}\ }\textbf {\bibinfo {volume} {75}},\
  \bibinfo {pages} {4031} (\bibinfo {year} {1995})}\BibitemShut {NoStop}%
\bibitem [{\citenamefont {Landau}\ and\ \citenamefont
  {Lifshitz}(1987)}]{landau1959fluid}%
  \BibitemOpen
  \bibfield  {author} {\bibinfo {author} {\bibfnamefont {Lev~Davidovich}\
  \bibnamefont {Landau}}\ and\ \bibinfo {author} {\bibfnamefont
  {Evgeny~Mikhailovich}\ \bibnamefont {Lifshitz}},\ }\href@noop {} \textit{
  {\bibinfo {title} {Fluid mechanics}}},\ \bibinfo {edition} {2nd}\ ed.,\
  \bibinfo {series} {Course of Theoretical Physics}, Vol.~\bibinfo {volume}
  {6}\ (\bibinfo  {publisher} {Butterworth--Heinemann},\ \bibinfo {address}
  {Oxford, UK},\ \bibinfo {year} {1987})\ \bibinfo {note} {an optional
  note}\BibitemShut {NoStop}%
\end{thebibliography}
%

\clearpage
\appendix
\onecolumngrid
\section{Supplemental Material}
\label{sec:supp}
\subsection{Model and Method}
\label{sec:MM}
Here we provide a detailed description of our model, the numerical
method and dimensionless parameters for the direct numerical
simulation. 
\subsubsection{Dynamical equations}
\label{sec:eqn}
We use a phase-field model for binary fluids with a Landau-Ginzburg type $\phi^4$ free energy.
The dynamics is that of a conserved order parameter coupled to an compressible
flow -- model H of Hohenberg and
Halperin~\cite{hoh+hal77} from which  the Langvein noise is excluded. 
\begin{subequations}
  \begin{align}
    (\delt + \vv\cdot\grad)\phi &= \Gamma \lap \mu \label{eq:phitA}\\
    \mu &= \frac{\delta \mF}{\delta \phi} \label{eq:muA}\\
    \mF[\phi] &= \frac{\Lambda}{\xi^2}\int d^dx\left[ f(\phi)
               + \frac{\xi^2}{2}\mid \grad\phi \mid^2 \right] \label{eq:LGA}\\
    f(\phi) &= \frac{1}{4}\left(1 - \phi^2 \right)^2 \label{eq:fenergyA}\\
    (\delt + \vv\cdot\grad)\vv &= \nu\lap\vv - \grad p -\phi\grad\mu
                       \label{eq:momA}\\
    \dive \vv &=0 \label{eq:divvA}
  \end{align}
  \label{eq:CHNSA}
\end{subequations}
As the density of the fluid is constant, we have set it to unity. 

\subsubsection{Dimensionless parameters}
We choose the radius of the droplets, $R$, as our characteristic length scale.
There is no characteristic velocity scale set by the initial condition, hence,
similar to what is done in convection~\cite{landau1959fluid}, we use
$\nu/R$, where $\nu$ is the viscosity, as our characteristic velocity scale
to obtain:
\begin{subequations}
  \begin{align}
    (\delt + \vv\cdot\grad)\phi &= \frac{1}{\Sc} \lap \mu\/, \label{eq:phi_ndim}\\
    (\delt + \vv\cdot\grad)\vv &= \lap\vv - \grad p -\frac{\Phiz^{1/3}\La}{\Ch}\phi\grad\mu 
    \label{eq:mom_ndim}
  \end{align}
  \label{eq:CHNS_ndim}
\end{subequations}
Here the four dimensionless numbers are:
the Cahn number, $\Ch \equiv \xi/\mL$,
the Schmidt number, $\Sc \equiv \nu/D$;
the Laplace number, $\La \equiv (\sigma R)/\nu^2$,
where $D \equiv \Gamma\Lambda/\xi^2$,
$\sigma = (2\sqrt{2}/3)\Lambda/\xi$
and the initial volume fraction occupied by the heavier phase
is $\Phiz$. 
A different non-dimensionalization~\cite{naso2018flow}
gives
Reynolds number $\Rey= (3/2\sqrt{2})^{1/2}(\mL/R)\sqrt{\La/\Ch}$
and 
Peclet number $\Pe = \Rey\Sc$.
\subsubsection{Numerical algorithm, initial condition, and parameters}
Our simulations are done in a cubic box of side $2\pi$.
  This box is discretized in $N$ equally spaced grid points in each direction.
  \Eq{eq:CH} is solved by using a pseudo-spectral method with one-half dealiasing and
  periodic boundary conditions.
  We use a second-order Adams-Bashforth scheme for time stepping.
  An earlier version of this code has been used before in
  Ref.~\cite{perlekar2014spinodal}.
  The velocity is zero at $t=0$.
  Initial conditions for order parameter $\phi$ are chosen to have an array of droplets as
  shown in \Fig{fig:pdiagram} (B), $\phi =1$ inside the drops and $-1$ outside.
Simulation parameters are listed in table~\ref{tab:param}. 

\begin{table}
  \caption{\label{tab:param} Parameters of simulations. The simulations are
    in $2\pi$ periodic domains in three-dimensions with $\tN^3$, $\tN=512$ and $1024$
    collocation points. All the simulations have $\Sc = 1$.
Initially, we choose all the droplets to have the same radius $R/\mL=1/(8\pi)$
and their centers placed on a cubic lattice.
We then add small random perturbations to the radii of
the droplets. 
The initial volume fraction occupied by the droplets (minority phase) is
approximately $0.2$ such that $\bra{\phi} = 0.62$.   
}
  \begin{ruledtabular}
    \begin{tabular}{llll}
      Runs & $\tN$ & $\Ch$ & $\La$ \\
\hline
      {\tt A0} &512 & $0.19 $ & $10^2$ \\
      {\tt A1} & 512 & $0.19 $ & $10^3$ \\
      {\tt A2} & 512 & $0.19 $ & $10^{4}$ \\
      {\tt A3} & 512& $0.19 $ & $10^{5}$ \\
      {\tt A4} & 512& $0.19 $ & $10^{6}$ \\
      {\tt A5} & 512& $0.05$ & $10^{6}$ \\
      {\tt A6} & 1024& $0.1$ & $10^{6}$ \\
    \end{tabular}
  \end{ruledtabular}
\end{table}

\subsubsection{Comparison of our parameters with earlier work}
\label{sec:comparison}
Recently Naso and Nar\'aigh~\cite{naso2018flow} have, for the first time,
ventured in to the part of the parameters space with
small diffusivisty and viscosity and found 
signatures of novel coarsening behavior in this regime.
Their simulations started with symmetric initial condition whereas
we start with droplets organized on a lattice with defects
 in the droplet-spinodal regime.

Nevertheless it is useful to compare the dimensionless parameters
of our simulations with theirs.
Ref.~\cite{naso2018flow} worked with dimensionless viscosity ($1/\Rey$) and
diffusivity ($1/\Pe$) and separated the parameters in different phases
depending on the types of structures that were observed, not based
on the dynamic scaling exponent. 
Our runs belong to the region on phase space Ref.~\cite{naso2018flow}
called ``anomalous'' and ``inertial''
\begin{figure}
  \includegraphics[width=0.5\textwidth]{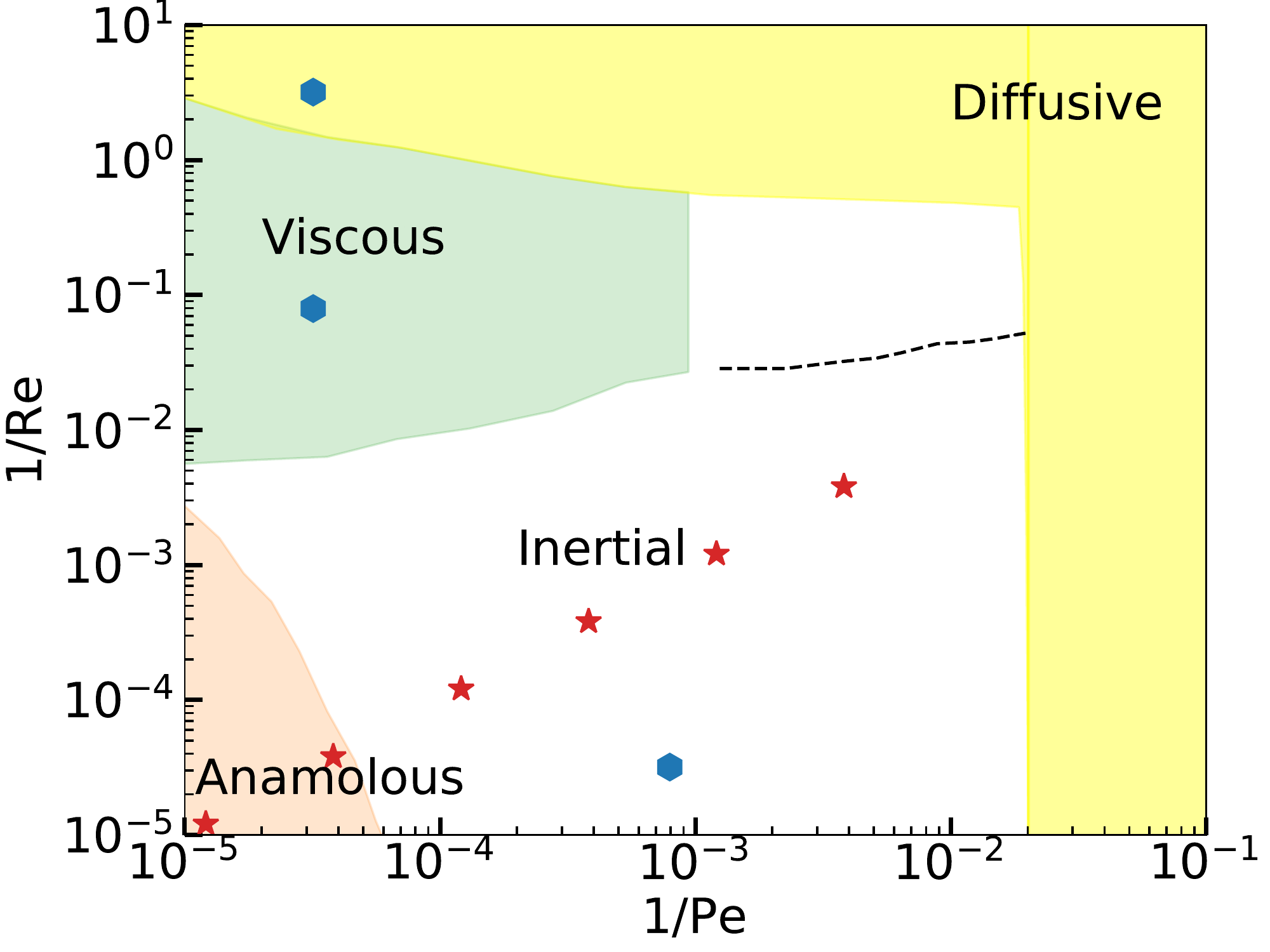}
  \caption{\label{fig:phase_plot} The phase diagram in the $1/\Pe$--$1/\Rey$
    plane according to Ref.~\cite{naso2018flow}. The points marked by red
    star are our simulations. We have also used our code to simulate
    the system with symmetric initial condition for the points marked in
    blue and have obtained results that agree with Ref.~\cite{naso2018flow}.
  }
\end{figure}
In the notation of Ref.~\cite{naso2018flow} the Cahn--Hilliard--Navier--Stokes
equations are:
\begin{align}
\rho D_t {\uu} &= \eta \nabla^2{\uu} -\nabla P - \phi \nabla \mu, \\
D_t \phi &= \frac{D}{\alpha} \nabla^2 \mu, \\
\mu &= \alpha(- \phi + \phi^3 - \xi^2 \nabla^2 \phi)~\rm{with}~\alpha\equiv \frac{\Lambda}{\xi^2}.
\label{eq:naso}
\end{align}

Choosing the velocity scale as  $U\equiv \sqrt{\alpha/\rho}$,
the length scale $\mL$ (box size) we obtain the following dimensionless
equations
\begin{align}
\rho D_t {\uu} &= \frac{1}{\Rey} \nabla^2{\uu} -\nabla P - \phi \nabla \mu, \\
D_t \phi &= \frac{1}{\Pe} \nabla^2 \mu, \\
\mu &= (- \phi + \phi^3 - \Ch^2 \nabla^2 \phi),
\label{eq:naso_ndim}
\end{align}
with $\Re\equiv U\mL/\nu$, $\Ch=\xi/\mL$ and $\Pe=U\mL/D$; the Reynolds
number, the Cahn number and Peclet number respectively, see
Table~\ref{tab:naso}.
\begin{table*}[!h]
  \caption{\label{tab:naso} Dimensionless numbers for our runs.
    Based on the values of $1/Pe$ and $1/Re$, we are in the
    anamolous regime of Ref.~\cite{naso2018flow}.}
\begin{ruledtabular}
\begin{tabular}{cccccccccc} 
Run & \multicolumn{1}{c} {$D$} & \multicolumn{1}{c} {$\Lambda$} & \multicolumn{1}{c} {$\xi$} & \multicolumn{1}{c} {$\nu$} &  \multicolumn{1}{c} {$\Ch$} & $\Pe$ & $\Rey$ \\
A0 & $5 \cdot 10^{-4}$ & $2.6 \cdot 10^{-2}$ & $2.45\cdot 10^{-2}$ & $5\cdot 10^{-4}$ &  $3.9\cdot 10^{-3}$ & $8.27 \cdot 10^{3}$ & $8.27\cdot 10^{3}$  \\
A1 & $5 \cdot 10^{-4}$ & $2.6 \cdot 10^{-3}$ & $2.45\cdot 10^{-2}$ & $5\cdot 10^{-4}$ &  $3.9\cdot 10^{-3}$ & $2.62 \cdot 10^{3}$ & $2.62\cdot 10^{3}$ \\
A2 & $5 \cdot 10^{-4}$ & $2.6 \cdot 10^{-4}$ & $2.45\cdot 10^{-2}$ & $5\cdot 10^{-4}$ &  $3.9\cdot 10^{-3}$ & $8.27 \cdot 10^{4}$ & $8.27\cdot 10^{4}$  \\
A3 & $5 \cdot 10^{-4}$ & $2.6 \cdot 10^{-5}$ & $2.45\cdot 10^{-2}$ & $5\cdot 10^{-4}$ &  $3.9\cdot 10^{-3}$ & $2.61 \cdot 10^{4}$ & $2.61\cdot 10^{4}$ \\
\end{tabular}
\end{ruledtabular}
\end{table*}
\subsection{Fluxes}
\label{sec:flux}
\begin{figure*}
	\includegraphics[width=0.48\textwidth]{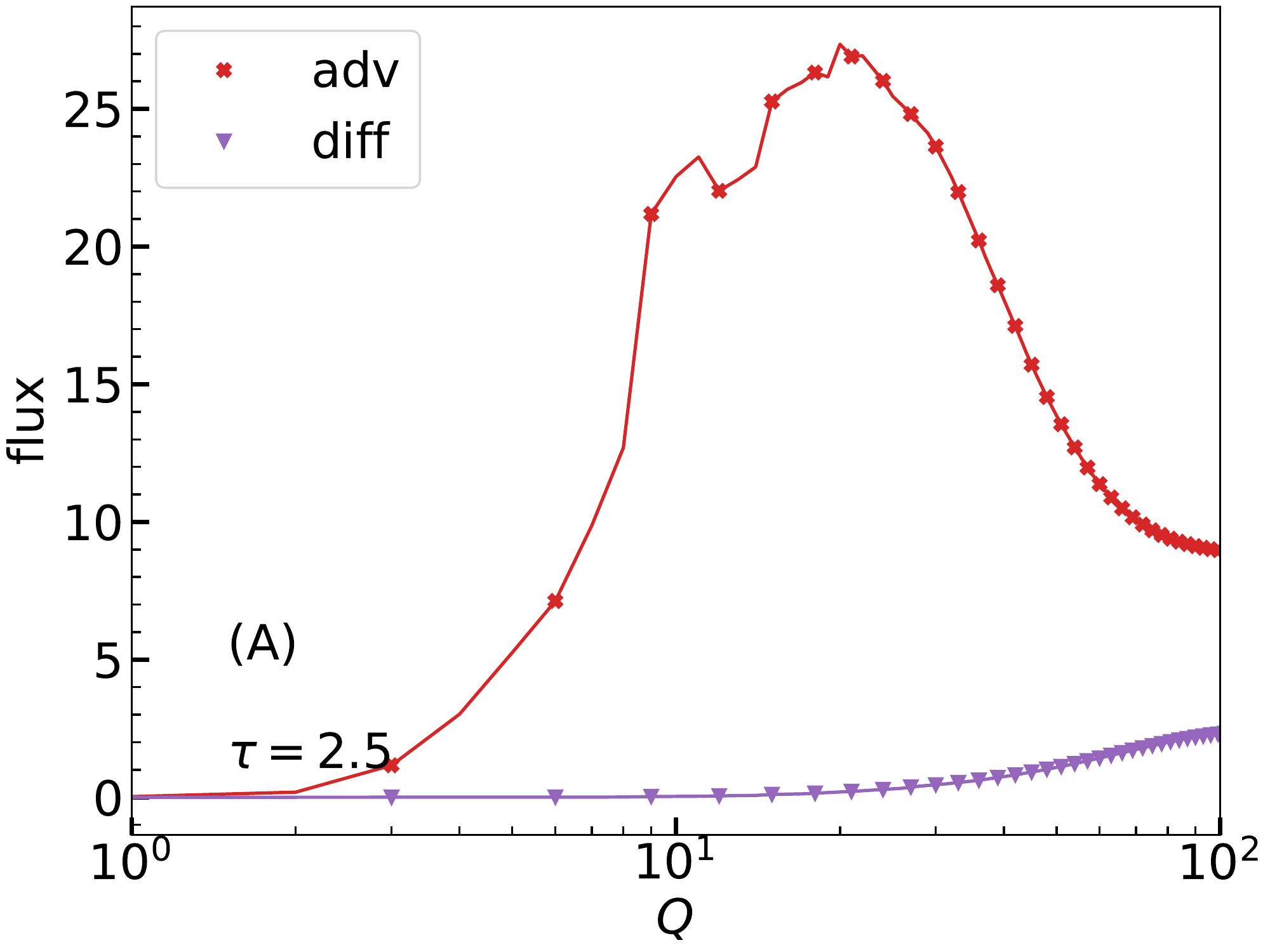}
	\includegraphics[width=0.48\textwidth]{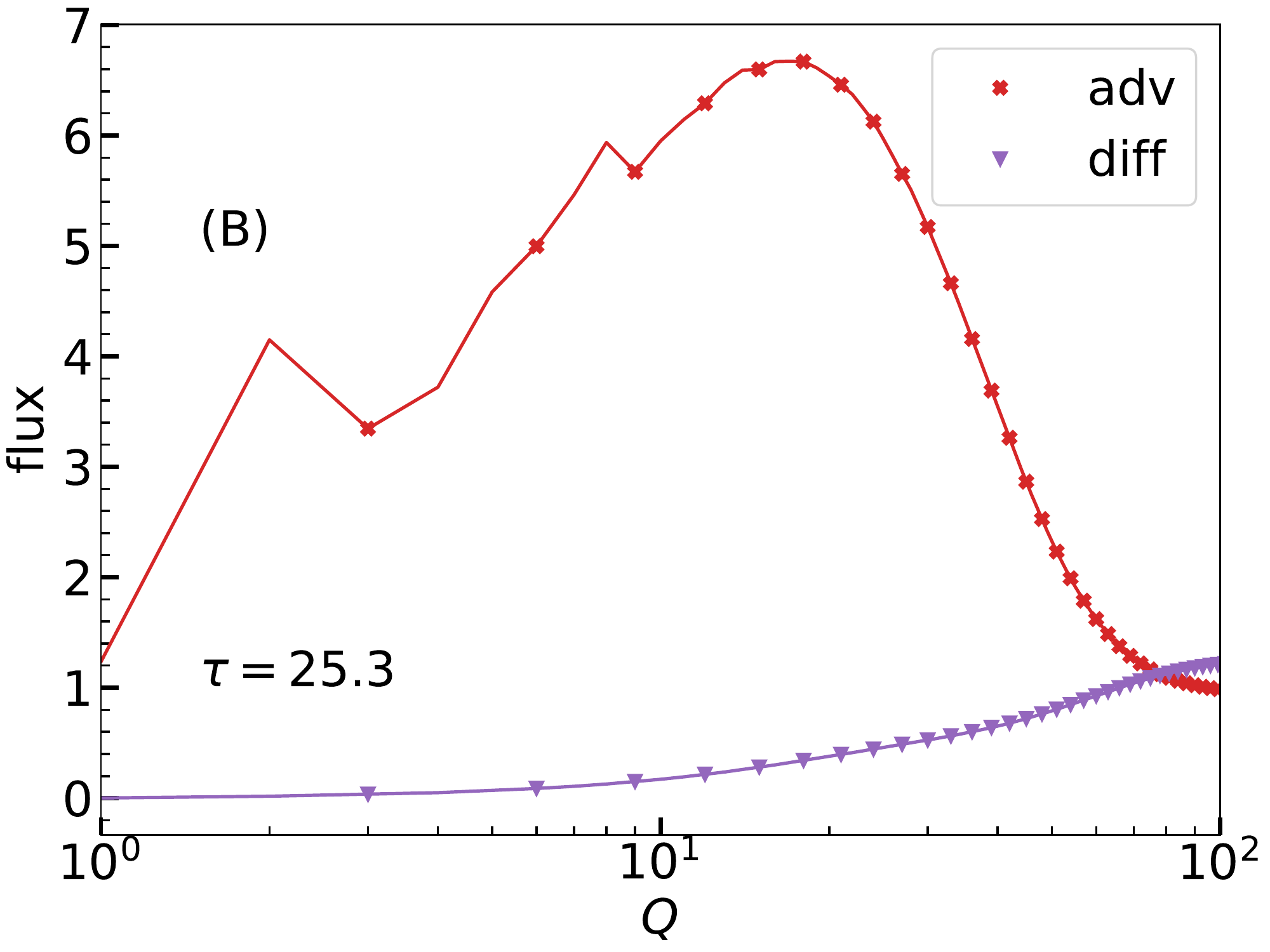}
  \caption{\label{fig:phi_flux} Contribution to the flux of $\phi$ from 
	the advective($\vv\cdot\grad\phi$) and the diffusive ($\Gamma\lap\mu$)
	term at early times (A) and intermediate times (B). 
    }
\end{figure*}

The standard approach to understand scaling behaviour in turbulence
is to calculate the scale-by-scale energy budget equation~\cite{Fri96}. 
Here we follow the procedure outline in \cite{perlekar2014spinodal}. 
Multiplying \eq{eq:phitA} by $\phat(-\qq)$ and
integrating over all the fourier modes upto wave-number $Q$,  we obtain
\begin{equation}
  \partial_t \SQ + \mAQ = \mGQ
  \label{eq:pflux}
\end{equation}
where $\mAQ$ is the contribution from the advective term and $\mGQ$ is the
contribution from the diffusive term.
In particular, 
\begin{subequations}
  \begin{align}
  \mAQ &\equiv \Re\left\{\int_0^Q \int_\Omega
  \phat(-\qq) \widehat{[\vv\cdot\grad\phi]}(\qq) dq d\Omega \right\} \/,
  \label{eq:AQ}\\
  \mGQ &\equiv \Re\left\{\Gamma \int_0^Q \int_\Omega  \phat(-\qq)
  \widehat{[\lap\mu]}(\qq)dq d\Omega \right\} \/.
  \label{eq:GQ}
  \end{align}
\end{subequations}
In \Fig{fig:phi_flux} we show two representative plots of these two fluxes
as a function of the wavenumber $Q$, one at short times and the other at
intermediate times.
We find that in all cases and at all $Q$, except for very high $Q$,
the advective flux dominates over the diffusive contribution. 

\begin{figure}
  \includegraphics[width=0.5\textwidth]{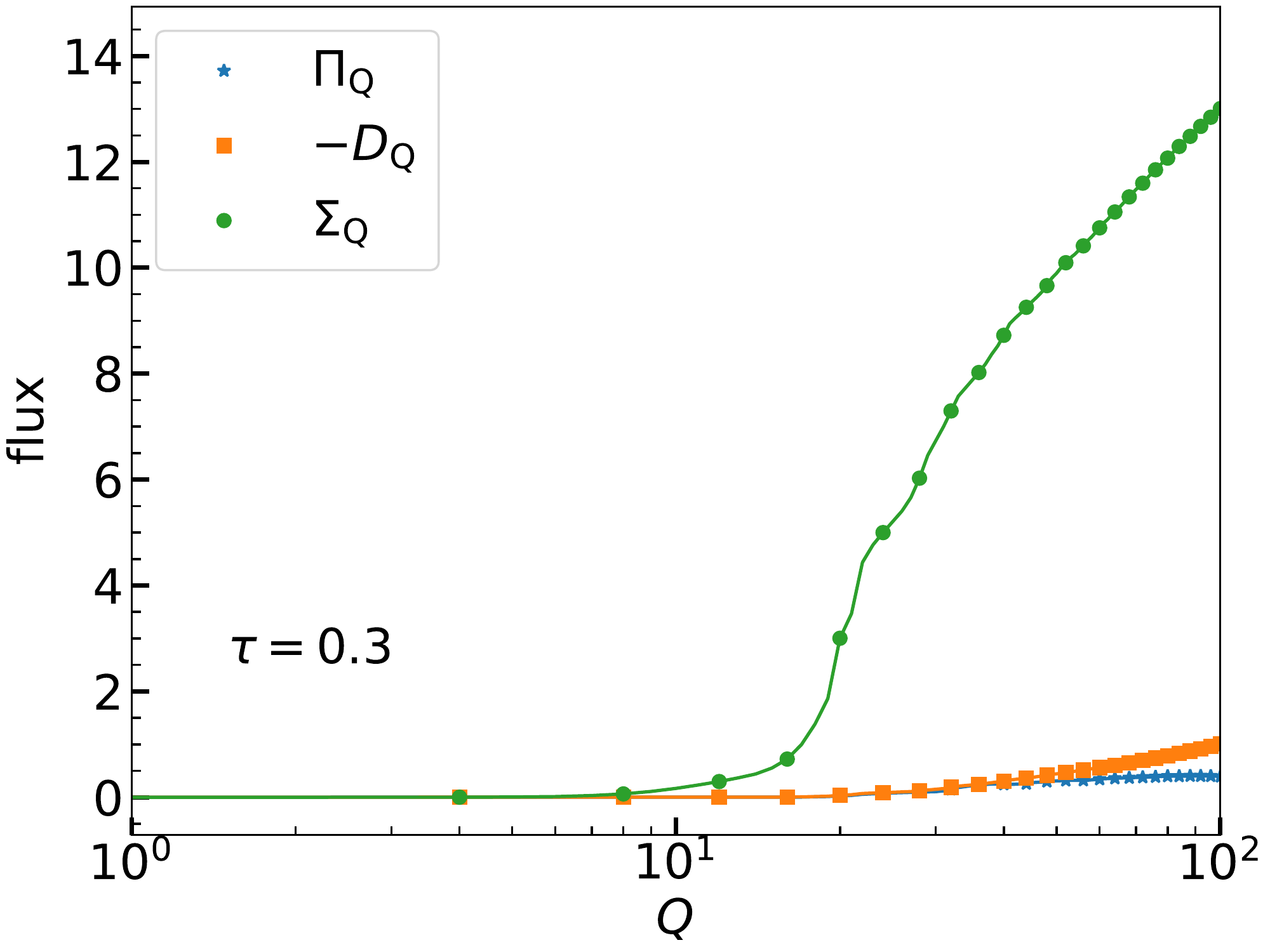}
  \caption{\label{fig:flux_shortt} Contribution to the flux of energy from
    the nonlinear term ($\PiQ$), the viscous term ($\DQ$) and 
    surface tension ($\SigmaQ$) at early time. 
  }
\end{figure}

By multiplying the Fourier transformed \Eq{eq:momA} with $\vhat{(-\qq)}$,
and integrating over all the fourier modes upto wave-number $Q$,  we obtain 
\begin{equation}
\partial_t \EQ = \PiQ + \DQ - \SigmaQ.
\label{eq:flux1}
\end{equation}
Here $\EQ$ is the cumulative kinetic energy.
The rest of the quantities in \eq{eq:flux1} are defined as follows:
\begin{subequations}
  \begin{align}
  \PiQ &\equiv -\Re\left\{\int_0^Q \int_\Omega
  \vhat(-\qq) \cdot \widehat{[\vv \cdot \grad \vv ]}(\qq) dq d\Omega \right\} 
  \label{eq:PiQ}\\
  \DQ &\equiv - \nu \int_0^Q  q^2 \bE(q) dq \label{eq:DQ} \\
  \SigmaQ &\equiv \Re\left\{ \int_0^Q \int_\Omega  \vhat(-\qq) \cdot
  \widehat{[\phi \nabla \mu]}(\qq)dq d\Omega \right\} 
  \end{align}
\end{subequations}
In \Fig{fig:flux_shortt} we show a representative plot of  all the three
contributions on the right hand side of \eq{eq:flux1} at short times.
Clearly $\SigmaQ$ dominates over the other two.
Hence we conclude that the system is not in a stationary state -- the
growth of kinetic energy is fuelled by the contribution from $\phi\grad\mu$. 
\clearpage
\twocolumngrid
\end{document}